\documentclass[twocolumn,floatfix]{aastex631}

\newcommand{\kms}{\hbox{km\,s$^{-1}$}}
\newcommand{\mjup}{$M_{\mathrm{Jup}}$}

\begin{document}
\title{CWISE J105512.11+544328.3: A Nearby Y Dwarf Spectroscopically Confirmed with Keck/NIRES}

\author[0009-0000-0638-6520]{Grady Robbins}
\affiliation{Department of Astronomy, University of Florida, 201 Criser Hall, Gainesville, FL 32611, USA}
\affiliation{NSF's National Optical-Infrared Astronomy Research Laboratory, 950 N. Cherry Ave., Tucson, AZ 85719, USA}

\author[0000-0002-1125-7384]{Aaron M. Meisner}
\affiliation{NSF's National Optical-Infrared Astronomy Research Laboratory, 950 N. Cherry Ave., Tucson, AZ 85719, USA}

\author[0000-0002-6294-5937]{Adam C. Schneider}
\affil{United States Naval Observatory, Flagstaff Station, 10391 West Naval Observatory Rd., Flagstaff, AZ 86005, USA}

\author[0000-0002-6523-9536]{Adam J.\ Burgasser}
\affiliation{Department of Astronomy \& Astrophysics, UC San Diego, 9500 Gilman Drive, La Jolla, CA 92093, USA}

\author[0000-0003-4269-260X]{J. Davy Kirkpatrick}
\affiliation{IPAC, Mail Code 100-22, California Institute of Technology, 1200 E. California Blvd., Pasadena, CA 91125, USA}

\author[0000-0002-2592-9612]{Jonathan Gagn\'e}
\affiliation{Plan\'etarium Rio Tinto Alcan, Espace pour la Vie, 4801 av. Pierre-de Coubertin, Montr\'eal, Qu\'ebec, Canada}
\affiliation{Institute for Research on Exoplanets, Universit\'e de Montr\'eal, D\'epartement de Physique, C.P.~6128 Succ. Centre-ville, Montr\'eal, QC H3C~3J7, Canada}

\author[0000-0002-5370-7494]{Chih-Chun Hsu}
\affiliation{Center for Interdisciplinary Exploration and Research in Astrophysics (CIERA), Northwestern University, 1800 Sherman, Evanston, IL, 60201, USA}

\author[0000-0001-7171-5538]{Leslie Moranta}
\affiliation{Plan\'etarium Rio Tinto Alcan, Espace pour la Vie, 4801 av. Pierre-de Coubertin, Montr\'eal, Qu\'ebec, Canada}
\affiliation{Institute for Research on Exoplanets, Universit\'e de Montr\'eal, D\'epartement de Physique, C.P.~6128 Succ. Centre-ville, Montr\'eal, QC H3C~3J7, Canada}

\author[0000-0003-2478-0120]{Sarah Casewell}
\affiliation{School of Physics and Astronomy, University of Leicester, University Road, Leicester, LE1 7RH, UK}

\author[0000-0001-7519-1700]{Federico Marocco}
\affiliation{IPAC, Mail Code 100-22, California Institute of Technology, 1200 E. California Blvd., Pasadena, CA 91125, USA}

\author[0000-0003-0398-639X]{Roman Gerasimov}
\affiliation{Center for Astrophysics and Space Science, University of California San Diego, La Jolla, CA 92093, USA}

\author[0000-0001-6251-0573]{Jacqueline K. Faherty}
\affil{Department of Astrophysics, American Museum of Natural History, Central Park West at 79th Street, NY 10024, USA}

\author[0000-0002-2387-5489]{Marc J. Kuchner}
\affil{Exoplanets and Stellar Astrophysics Laboratory, NASA Goddard Space Flight Center, 8800 Greenbelt Road, Greenbelt, MD 20771, USA}

\author[0000-0001-7896-5791]{Dan Caselden}
\affil{Department of Astrophysics, American Museum of Natural History, Central Park West at 79th Street, NY 10024, USA}

\author[0000-0001-7780-3352]{Michael C. Cushing}
\affil{Ritter Astrophysical Research Center, Department of Physics \& Astronomy, University of Toledo, 2801 W. Bancroft St., Toledo, OH 43606, USA}

\author[0000-0003-0548-0093]{Sherelyn Alejandro}
\affiliation{Hunter College, City University of New York, 695 Park Avenue, NY 10065, USA}
\affil{Department of Astrophysics, American Museum of Natural History, Central Park West at 79th Street, NY 10024, USA}

\author{The Backyard Worlds: Planet 9 Collaboration}

\author{The Backyard Worlds: Cool Neighbors Collaboration}

\begin{abstract}

Y dwarfs, the coolest known spectral class of brown dwarfs, overlap in mass and temperature with giant exoplanets, providing unique laboratories for studying low-temperature atmospheres. However, only a fraction of Y dwarf candidates have been spectroscopically confirmed. We present Keck/NIRES near-infrared spectroscopy of the nearby ($d \approx 6-8$ pc) brown dwarf CWISE J105512.11+544328.3.  Although its near-infrared spectrum aligns best with the Y0 standard in the $J$-band, no standard matches well across the full $YJHK$ wavelength range. The CWISE J105512.11+544328.3 NH$_3$-$H$ = 0.427 $\pm$ 0.0012 and CH$_4$-$J$ = 0.0385 $\pm$ 0.0007 absorption indices and absolute Spitzer [4.5] magnitude of 15.18 $\pm$ 0.22 are also indicative of an early Y dwarf rather than a late T dwarf. CWISE J105512.11+544328.3 additionally exhibits the bluest Spitzer [3.6]$-$[4.5] color among all spectroscopically confirmed Y dwarfs. Despite this anomalously blue Spitzer color given its low luminosity, CWISE J105512.11+544328.3 does not show other clear kinematic or spectral indications of low metallicity. Atmospheric model comparisons yield a log(g) $\le$ 4.5 and $T_{\rm eff} \approx 500 \pm 150$~K for this source. We classify CWISE J105512.11+544328.3 as a Y0 (pec) dwarf, adding to the remarkable diversity of the Y-type population. JWST spectroscopy would be crucial to understanding the origin of this Y dwarf's unusual preference for low-gravity models and blue 3-5~$\mu$m color.
\end{abstract}


\section{Introduction} \label{sec:intro}

Over the past three decades, the development of increasingly sensitive infrared telescopes and detectors \citep[e.g.,][]{Rieke_review} has correspondingly led to the discovery of cooler, lower luminosity classes of substellar objects. Most recently, NASA's Wide-field Infrared Survey Explorer \citep[WISE;][]{Wright_2010} has revealed the Y-type spectral class \citep[$T_{\rm eff} \lesssim 500$ K;][]{Cushing_2011, Kirkpatrick_2011}, which extends to temperatures at least as low as $\approx 250$ K \citep{Luhman_2014}.

Y dwarfs are particularly important in that their masses and temperatures overlap with those of giant exoplanets, providing ideal laboratories for studying atmospheric chemistry without the glare of a primary star \citep[e.g.,][]{Leggett_2019}. However, there appears to be significant diversity among the Y dwarfs \citep[e.g.,][]{Beichman_2013,Leggett_2017,Miles_2020,Leggett_2021,Kirkpatrick_2021,Faherty_JWST}. One manifestation of this diversity is the relatively large spread in absolute magnitudes near the spectral energy distribution's peak ($\lambda \sim 4.5$~$\mu$m) at fixed 3$-$5~$\mu$m color (see e.g., Figure 18c of \citealt{Kirkpatrick_2021}). Whether this diversity arises from fundamental differences in formation scenario (e.g., gravitational collapse of a molecular cloud versus exoplanet ejection), large sensitivity to atmospheric abundances (i.e., metallicity), cloud properties, viewing geometry \citep[e.g.,][]{Vos_2017}, or multiplicity remains unknown. The diversity of Y dwarfs is perhaps not entirely surprising in light of the varied properties seen among our own solar system's giant planets \citep[e.g.,][]{Guillot_review}. JWST mid-infrared spectroscopy may rewrite our understanding of Y dwarfs and the T/Y boundary, but more examples and spectra of Y dwarfs are needed.

Here we present new near-infrared spectroscopy of the brown dwarf CWISE J105512.11+544328.3 (hereafter W1055+5443). Though originally thought to be a T8 dwarf based on its Spitzer color, a recent parallax measurement \citep{Kirkpatrick_2021} has placed this object much closer to the Sun than previously anticipated, at a distance of only $\sim 7$ pc. This nearby distance implies a very low luminosity consistent with that of a Y-type brown dwarf \citep{Kirkpatrick_2021}. In this work, we compare our Keck/NIRES spectrum against spectral standards, confirming a Y dwarf spectral classification for W1055+5443 in the near-infrared. The W1055+5443 near-infrared spectrum is anomalous, with a $J$-band 
 peak characteristic of a Y0 dwarf but a $K$-band morphology reminiscent of low gravity models and/or higher temperature. W1055+5443 thus adds to our evolving picture of Y dwarf diversity.

In Section \ref{sec:observe} we describe our Keck/NIRES spectroscopic observations and W1055+5443 $J$-band photometric detection newly extracted from archival imaging. In Section \ref{sec:spec analysis} we present our analysis of the W1055+5443 near-infrared spectrum, including the determination of its spectral type and atmospheric properties through brown dwarf standard comparisons, model fitting, and measurements of spectral indices. Finally, in Section \ref{sec:discussion} we synthesize our findings.

\section{Archival Data \& New Observations}\label{sec:observe}

W1055+5443 was initially identified as a WISE galaxy or dwarf candidate by \cite{2012AJ....144..148G} and was later confirmed via its proper motion to be a nearby brown dwarf \citep{Kirkpatrick_2021,Backyard_Worlds}. \cite{2012AJ....144..148G} measured a Spitzer ch1$-$ch2 color\footnote{The Spitzer ch1 bandpass has a central wavelength of 3.6~$\mu$m and the Spitzer ch2 bandpass has a central wavelength of 4.5~$\mu$m.} of 1.84 $\pm$ 0.04 mag and \cite{Kirkpatrick_2021} obtained a Spitzer-based parallax placing W1055+5443 surprisingly close to the Sun ($\varpi_{abs}$ = 145.0 $\pm$ 14.7 mas; $d$ = 6.9$_{-0.6}^{+0.8}$ pc). Table \ref{tab:properties} summarizes the relevant properties and photometry of W1055+5443 discussed throughout this paper. All magnitudes quoted throughout this work are in the Vega system unless specifically noted otherwise. All of the CatWISE2020 \citep{Marocco_2021} $W1$ and $W2$ magnitudes quoted in this work use the `mpro$\_$pm' columns.

\begin{table}[htbp]
\raggedright
\caption{CWISE J105512.11+544328.3 Properties and Photometry}
\label{tab:properties}
\begin{tabular}{lll}
\hline\hline
\multicolumn{3}{c}{Properties} \\
\hline\hline
Parameter & Value & References \\
\hline
$\mu_\alpha$ (mas yr$^{-1}$) & $-$1518.7 $\pm$ 2.1 & 2 \\
$\mu_\delta$ (mas yr$^{-1}$) &  $-$222.7 $\pm$ 2.0 & 2 \\
$\mu_{tot}$ (mas yr$^{-1}$) & 1534.9 $\pm$ 2.9 & 2 \\
$\varpi_{\rm abs}$ (mas) & 145.0 $\pm$ 14.7 & 2 \\
$v_{\rm tan}$ (km s$^{-1}$) & 50.2 $\pm$ 5.2 & 1 \\
Sp. Type & Y0 (pec) $\pm$ 0.5 & 1 \\
$T_{\rm eff}$ (K) & 500 $\pm$ 150  & 1 \\
log(g) (cgs)  & $\leq 4.5$   & 1 \\
\hline\hline
\multicolumn{3}{c}{Photometry} \\
\hline\hline
Parameter & Value & References \\
\hline
$J_{\rm MKO}$ (mag) & 18.868 $\pm$ 0.066 & 1 \\
$J_{\rm 2MASS}$ (mag) & $> 18.84$   & 2 \\
$H$ (mag) & $> 18.02$  & 2 \\
$Ks$ (mag) & $> 16.81$  & 2 \\
CatWISE2020 $W1$ (mag) & 17.332 $\pm$ 0.082 & 4 \\
CatWISE2020 $W2$ (mag) & 14.371 $\pm$ 0.018 & 4 \\
AllWISE $W1$ (mag) & 17.306 $\pm$ 0.127 & 3 \\
AllWISE $W2$ (mag) & 14.366 $\pm$ 0.044 & 3 \\
AllWISE $W3$ (mag) & 11.553 $\pm$ 0.196 & 3 \\
Spitzer ch1 (mag) & 16.219 $\pm$ 0.033 & 2 \\
Spitzer ch2 (mag) & 14.376 $\pm$ 0.019 & 2 \\
\hline
\end{tabular} \\[1ex]
\footnotesize \textbf{References}— (1) This work (2) \citealt{Kirkpatrick_2021} (3) \citealt{2013wise.rept....1C} (4) \citealt{Marocco_2021}.
\end{table}

\begin{figure}[ht]
\includegraphics[scale=0.47]{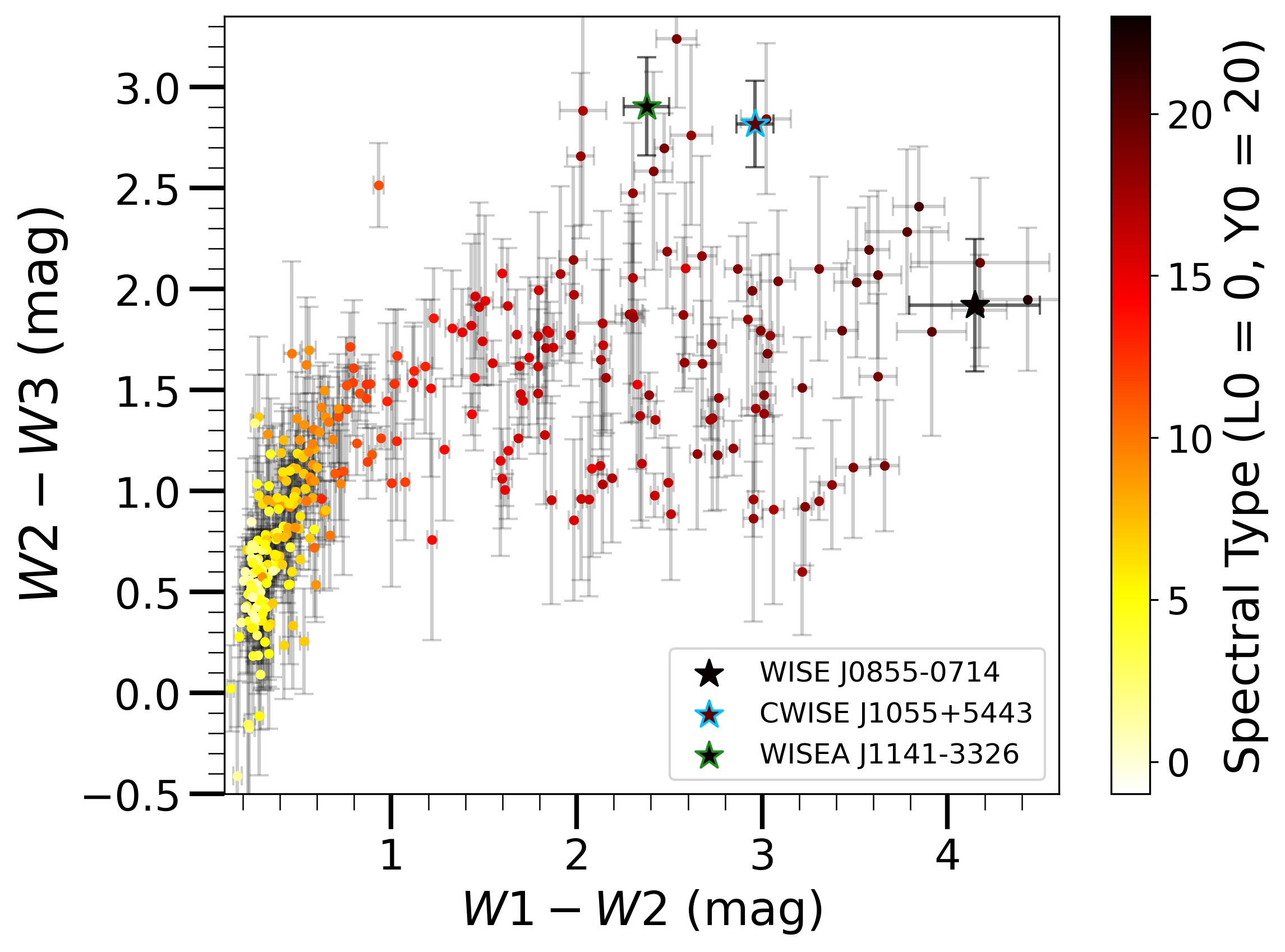}
\caption{\label{fig:W23vW12}
CatWISE2020 and AllWISE $W2$$-$$W3$ color versus CatWISE2020 $W1$$-$$W2$ color for the LTY dwarfs in \cite{Kirkpatrick_2021} with CatWISE detections in $W1$ and $W2$ and AllWISE detections in $W3$. The blue star is W1055+5443, the green star is Y0 dwarf WISEA J114156.67$-$332635.5, and the black star is Y dwarf WISE 0855$-$0714 \citep{Luhman_2014}. The WISE 0855$-$0714 data point makes use of this object's $W3$ detection reported in \cite{Leggett_2017}. Note that the $W1$ photometry of WISEA J114156.67$-$332635.5 is likely contaminated (resulting in an artificially blue $W1$$-$$W2$ color), given that the AllWISE epoch and Spitzer ch1 epoch for this source are similar (see Section \ref{subsec:spitzer_color} for discussion of the Spitzer ch1 contamination of this source).}
\end{figure}

To contextualize W1055+5443 within the population of known brown dwarfs, we generated multiple color-color, color-type, and color-magnitude diagrams (Figures \ref{fig:W23vW12}-\ref{fig:ch12J}). Figure \ref{fig:W23vW12} is a scatter plot of $W2$$-$$W3$ color against $W1$$-$$W2$ color, where the photometry is from CatWISE2020 ($W1$, $W2$), and AllWISE $W3$. Figure \ref{fig:W23vW12} shows that W1055+5443 has one of the reddest $W2$$-$$W3$ colors among known brown dwarfs detected in all of $W1$, $W2$, and $W3$. 

\subsection{Spitzer Photometry} \label{subsec:spitzer_color}

\begin{figure}[ht!]
\includegraphics[scale=0.47]{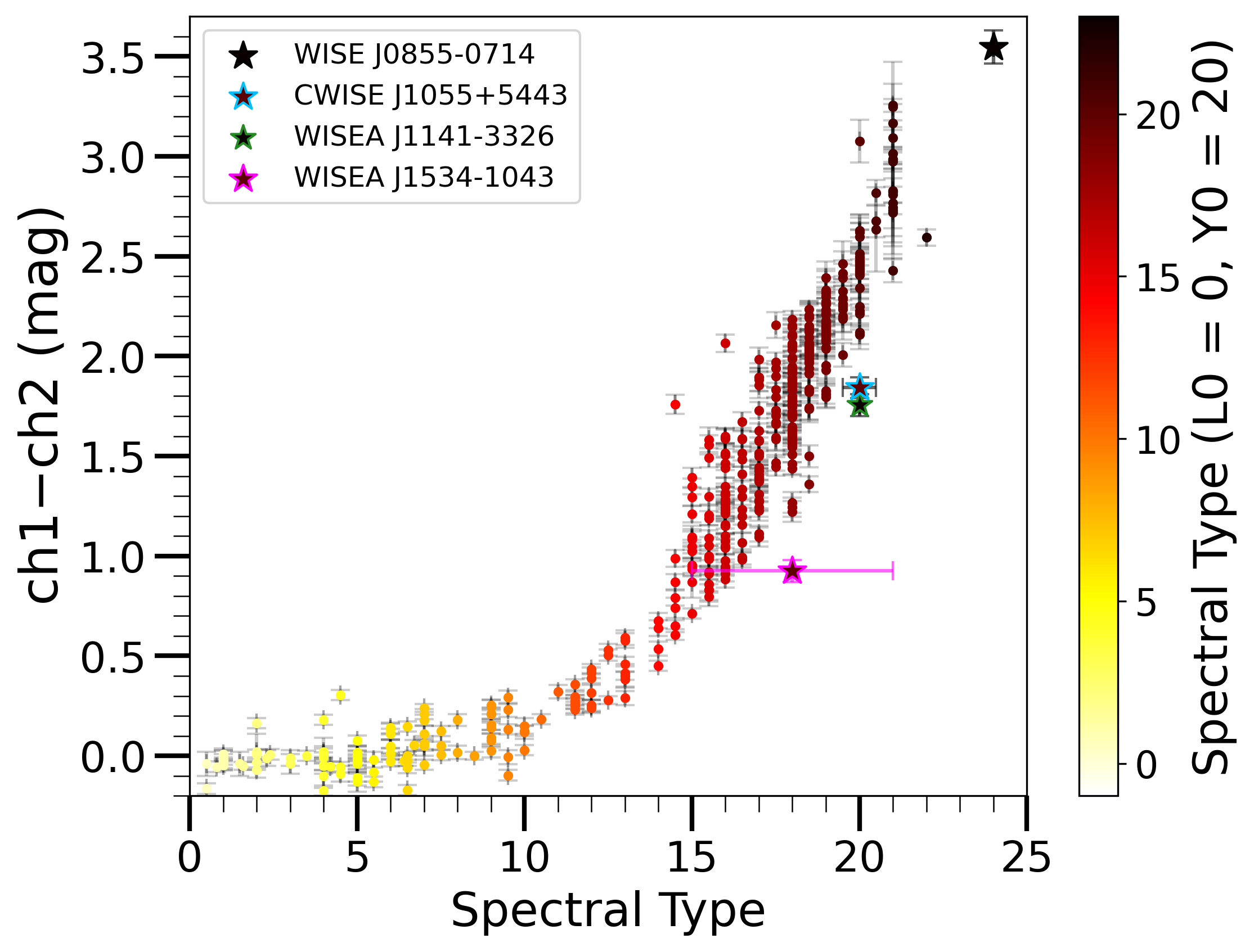}
\caption{\label{fig:ch12Sp}
Spitzer ch1$-$ch2 color plotted against spectral type for the LTY dwarfs in \cite{Kirkpatrick_2021}, where the blue star is W1055+5443, the purple star is WISEA J153429.75$-$104303.3 \citep[a.k.a. The Accident;][]{CatWISE_Spitzer,The_Accident}, the green star is WISEA J114156.67$-$332635.5, and the black star is Y dwarf WISE 0855$-$0714. Cases of objects with Spitzer limits rather than detections in ch1 and/or ch2 are excluded. See Section \ref{sec:spec analysis} for our spectral typing of W1055+5443.}
\end{figure}

W1055+5443 was observed with Spitzer Space Telescope's Infrared Array Camera (IRAC) instrument \citep{Spitzer_Space_Telescope,IRAC} on multiple occasions, ranging from 2011 January to 2019 September (PI Kirkpatrick; PID 70062, 14224). There are many archival epochs of Spitzer/IRAC 4.5~$\mu$m (ch2 a.k.a. [4.5]) imaging available, but only one epoch of Spitzer/IRAC 3.6~$\mu$m (ch1 a.k.a. [3.6]) imaging available. The absolute ch2 magnitude of W1055+5443 ($M_{\rm ch2} = $15.18 $\pm$ 0.22) corresponds to a spectral type of Y0 to Y1 according to the polynomial relations of \cite{Kirkpatrick_2021}. However, the W1055+5443 Spitzer color of ch1$-$ch2 = 1.84 $\pm$ 0.04 mag is exceptionally blue compared to any other Y dwarf. Among the Y0 or later dwarfs tabulated in \cite{Kirkpatrick_2021}, only WISEA J114156.67$-$332635.5 has a slightly bluer color, with ch1$-$ch2 = 1.755 $\pm$ 0.041 mag (see Figure \ref{fig:ch12Sp}). However, WISEA J114156.67$-$332635.5's peculiar ch1$-$ch2 color is attributed to contamination from a background source at its Spitzer ch1 observation epoch \citep{Kirkpatrick_2019,Kirkpatrick_2021}. Thus, W1055+5443 has the bluest Spitzer ch1$-$ch2 color among spectroscopically confirmed Y dwarfs\footnote{We note that WISEA J153429.75$-$104303.3 \citep[a.k.a. The Accident;][]{CatWISE_Spitzer,The_Accident}, currently thought to be a metal-poor halo dwarf near the T/Y boundary, has a bluer Spitzer ch1$-$ch2 color than W1055+5443. However, no spectrum of WISEA J153429.75$-$104303.3 is available and its temperature remains uncertain.}. 

To verify that the Spitzer [3.6] and [4.5] magnitudes for W1055+5443 are accurate, we scrutinized the Spitzer images of W1055+5443 and its surrounding sky region. There were seven ch2 imaging epochs during 2019, allowing us to view the sky location where W1055+5443 was previously located during the 2011 ch1 imaging epoch; no contaminating source is visible at that location in the 2019 ch2 data. There is only a single imaging epoch available in ch1, comprised of five individual exposures. In the third of five ch1 exposures, there is a cosmic ray nearby the W1055+5443 location, but this was removed via outlier rejection during construction of the mosaic used for ch1 photometry \citep{Kirkpatrick_2021}. Any contaminant at the 2011 position of W1055+5443 would need to be an object which is not visible in any of the 2019 ch2 epochs or the relevant PanSTARRS1 images \citep{PS1}, yet has significant ch1 flux. This scenario is implausible, so we conclude that the blue Spitzer ch1$-$ch2 color for W1055+5443 is accurate. As a further crosscheck on the W1055+5443 Spitzer ch1 magnitude, we used the \cite{Kirkpatrick_2021} polynomials to predict a $M_{\rm ch1}$ value from our CatWISE2020 $M_{W1}$ measurement, which results in a prediction of $M_{\rm ch1}$ = 17.058 $\pm$ 0.387. This is well within the uncertainty of our actual $M_{\rm ch1}$ = 17.026 $\pm$ 0.221 measurement, and we therefore conclude that the Spitzer ch1 magnitude is consistent with the CatWISE2020 $W1$ and AllWISE $W1$ magnitudes.

\begin{figure}[ht]
\includegraphics[scale=0.47]{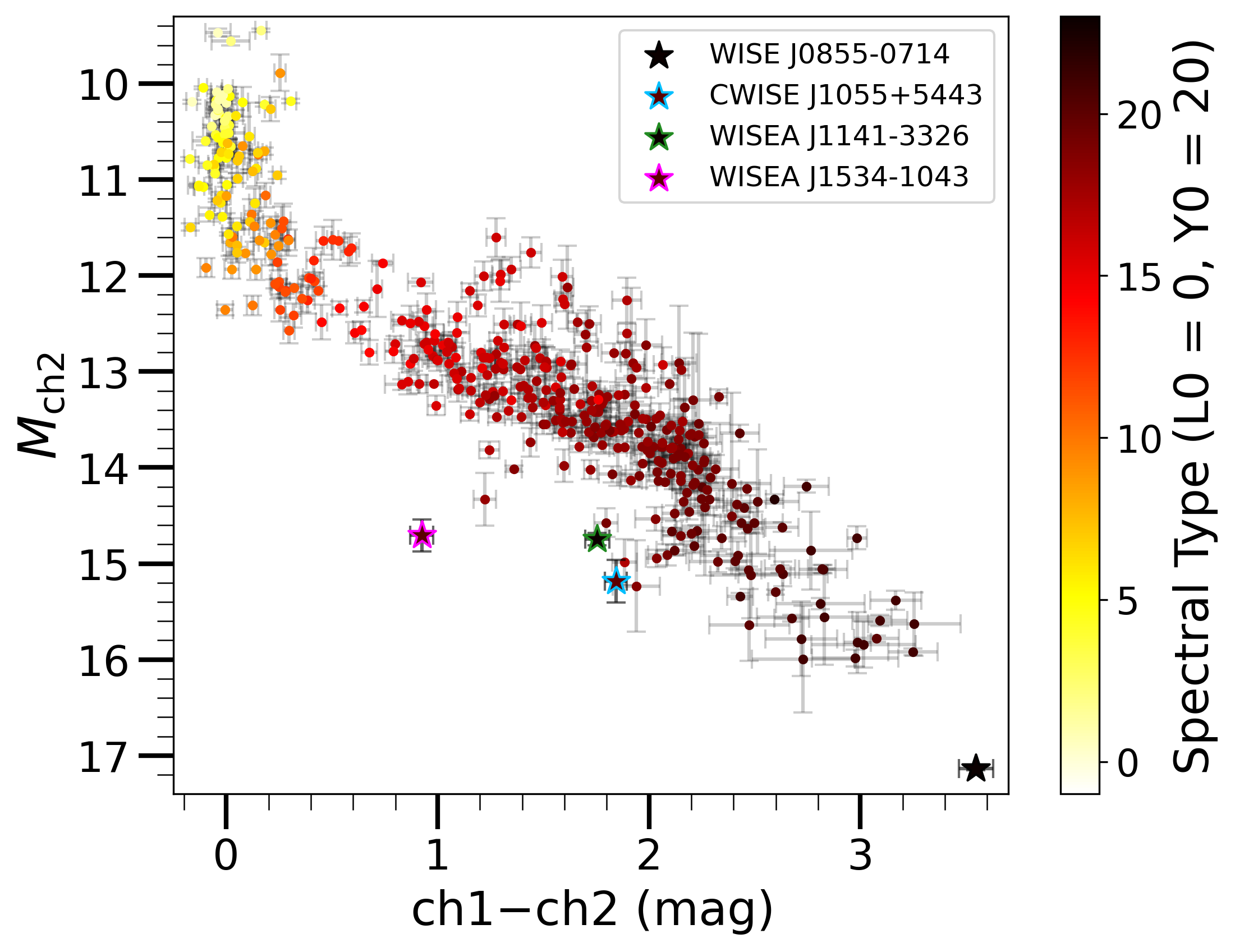}
\caption{\label{fig:ch12vMch2}
Spitzer ch2 absolute magnitude ($M_{\rm ch2}$) plotted against Spitzer ch1$-$ch2 color for the LTY dwarfs in \cite{Kirkpatrick_2021}, where the blue star is W1055+5443, the purple star is WISEA J153429.75$-$104303.3 (a.k.a. The Accident), the green star is WISEA J114156.67$-$332635.5, and the black star is Y dwarf WISE 0855$-$0714}. Cases of objects with Spitzer limits rather than detections in ch1 and/or ch2 are excluded, as are dwarfs without parallax measurements available.
\end{figure}

Using the ch1 and ch2 photometry, we can make several additional Spitzer-based comparisons of W1055+5443 against the population of brown dwarfs that also have ch1 and ch2 photometric detections available. Figure \ref{fig:ch12vMch2} shows ch2 absolute magnitude versus ch1$-$ch2 color. The combination of a relatively blue ch1$-$ch2 color coupled with a faint $M_{\rm ch2}$ renders W1055+5443 an outlier compared to the overall brown dwarf locus, with its characteristics most closely corresponding to those of late T and early Y dwarfs. The broadband spectral energy distribution of W1055+5443 is unusual, casting doubt on any color-based photometric type estimates and underscoring the importance of spectroscopic classification (Section \ref{sec:spec analysis}).

\subsection{J-band Photometry} \label{subsec:photometry}

The area of sky containing W1055+5443 was imaged twice as part of the UKIRT Hemisphere Survey (UHS; \citealt{Dye_2017}), once in 2016 and again in 2020.  However, both of these frames were deprecated, and therefore $J$-band photometry of W1055+5443 was not included in either the UHS DR1 or DR2 catalog releases. Calibrated photometry for these observations was found through the WFCAM Science Archive in the supplementary UHSDetectionAll table, which contains extracted photometry for all UHS pawprints including deprecated detections. We found $J$-band Vega magnitudes of 18.776 $\pm$ 0.105 mag and 18.929 $\pm$ 0.085 mag for the 2016 and 2020 epochs, respectively. To obtain a final $J$ magnitude for W1055+5443, we combined the 2016 and 2020 detection magnitudes using inverse-variance weighting, resulting in $J$ = 18.868 $\pm$ 0.066 mag, which we report in Table \ref{tab:properties}.

\begin{figure}[ht]
\includegraphics[scale=0.47]{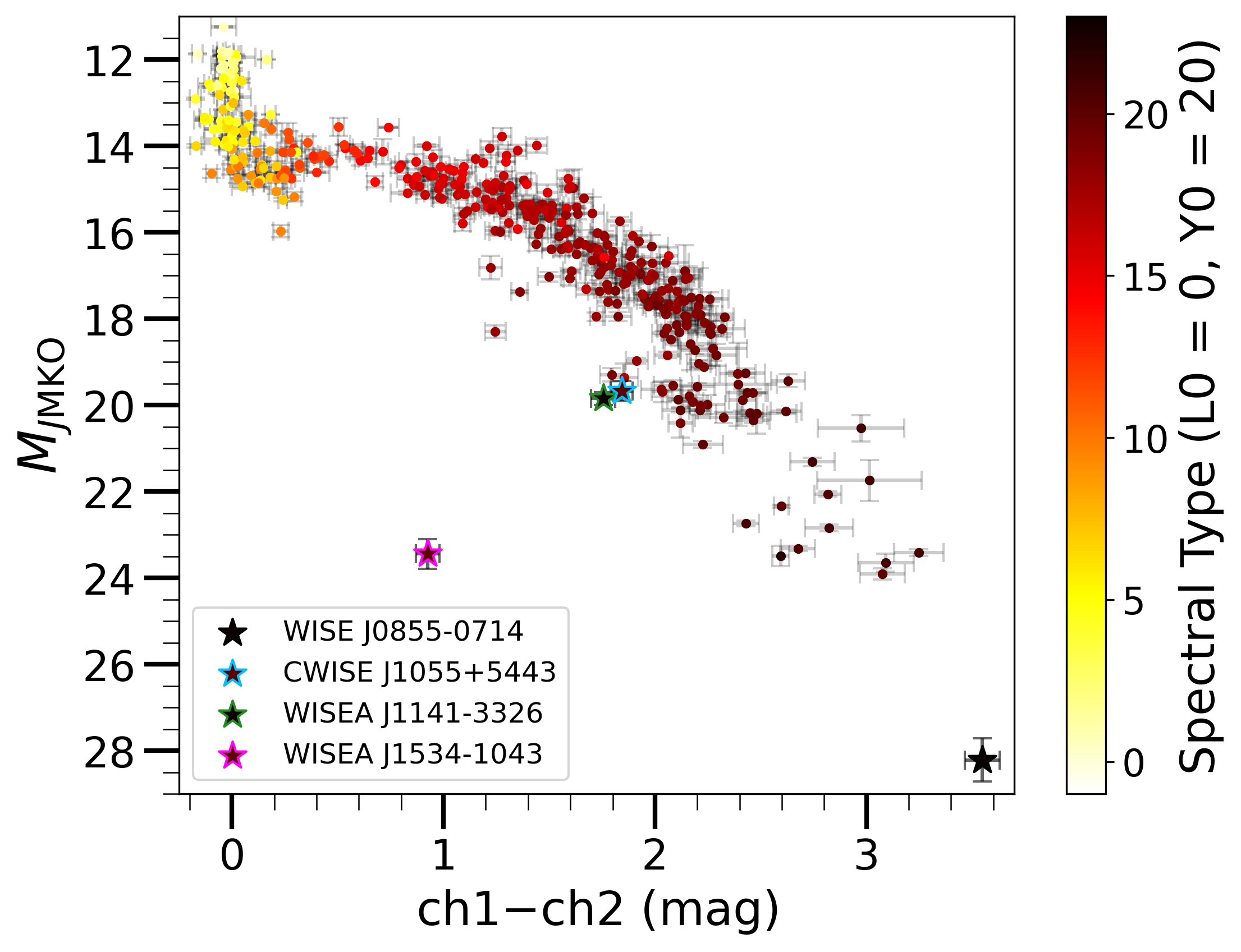}
\caption{\label{fig:ch12J}
$J$-band absolute magnitude ($M_{J\rm MKO}$) plotted against Spitzer ch1$-$ch2 color for the LTY dwarfs in \cite{Kirkpatrick_2021}, where the blue star is W1055+5443, the purple star is WISEA J153429.75$-$104303.3 (a.k.a. The Accident), the green star is WISEA J114156.67$-$332635.5, and the black star is Y dwarf WISE 0855$-$0714. Cases of objects with Spitzer and/or $J$-band limits rather than detections are excluded, as are dwarfs without parallax measurements available. The WISEA J153429.75$-$104303.3 data point makes use of its Gemini $J$-band detection from \cite{Meisner_2023}.}
\end{figure}

With our UHS-based $J$-band magnitude for W1055+5443 in hand, we can plot a color-magnitude diagram of absolute $J$-band magnitude ($M_{J \rm MKO}$) versus ch1$-$ch2 color (Figure \ref{fig:ch12J}). Due to its combination of faint $M_{J \rm MKO}$ and relatively blue ch1$-$ch2 color, W1055+5443 deviates significantly from the brown dwarf locus. The $M_{J \rm MKO}$ = 19.67 $\pm$ 0.25 mag value for W1055+5443 would typically correspond to a spectral type of T9.5-Y0 \citep{Kirkpatrick_2021}. Our $J$-band apparent magnitude implies a $J$$-$ch2 color of 4.49 $\pm$ 0.07 mag for W1055+5443, which would typically correspond to a spectral type of $\approx$~T9-T9.5 \citep{Kirkpatrick_2021}.

\subsection{Near-infrared Spectroscopy} \label{subsec:infraspec}

We used the Near-Infrared Echellette Spectrometer (NIRES; \citealt{2004SPIE.5492.1295W}) on the Keck II telescope on 2021 February 24 (UT) to obtain 0.94–2.45 $\mu$m near-infrared spectra of W1055+5443. Keck/NIRES has a fixed instrument configuration with a 0\farcs55 slit producing spectral resolution $\lambda/\Delta\lambda \sim$ 2700. 
Conditions were clear with a seeing of 0$\farcs$5.
The target was visible in the $K$-band slit-viewing camera and placed into the spectroscopic slit that was aligned with the parallactic angle. We obtained a set of 5 $\times$ 300~second exposures in an ABABB nodding pattern along the slit over an airmass range of 1.2–1.3. The A0 V star HD 56385 (V = 8.1 mag) was observed immediately afterward for flux calibration and telluric correction, and flat-field lamp exposures were obtained for pixel response calibration. 

\begin{figure*}[ht]
\includegraphics[scale=0.68]{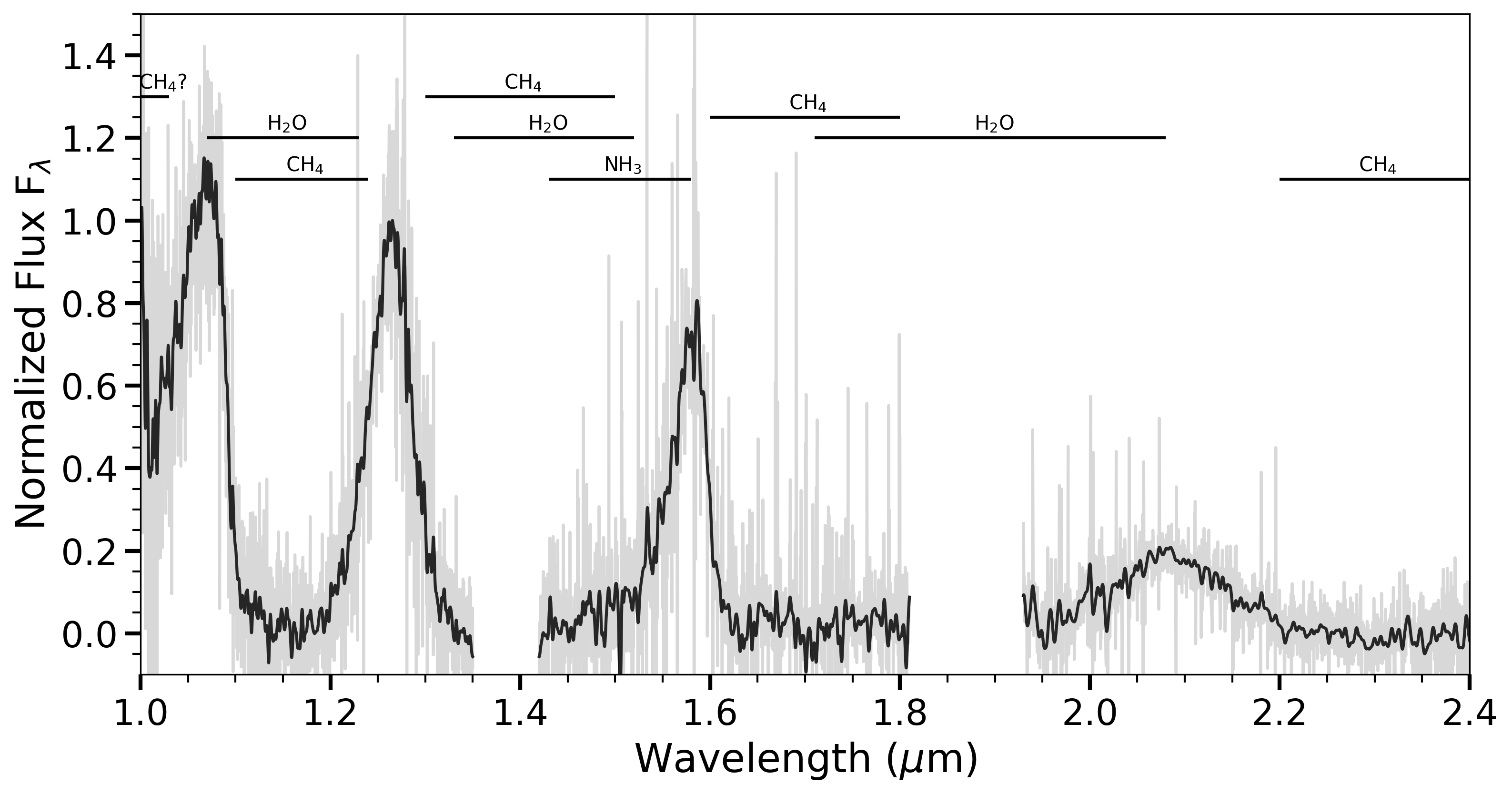}
\caption{\label{fig:IndivSpectra}
Keck/NIRES near-infrared spectrum for CWISE J105512.11+544328.3: original spectrum (gray) and 5-pixel smoothed spectrum (black). The spectrum is normalized to the $J$-band peak between 1.27 and 1.29 $\mu$m.}
\end{figure*}

Data were reduced using a modified version of Spextool \citep{Vacca_2003,Cushing_2004} following the standard procedure, which includes: pixel response calibration, wavelength calibration using sky OH emission lines, and spatial and spectral rectification using flat-field and telluric line exposures; optimal extraction of individual spectra from A-B pairwise subtracted frames; combination of these spectra with outlier masking; and telluric correction using the A0 V spectrum following \cite{Vacca_2003}. The reduced Keck/NIRES spectrum smoothed to $\lambda/\Delta\lambda \sim$ 500 is shown in Figure \ref{fig:IndivSpectra}, with further comparisons to standards and atmospheric models in Figures \ref{fig:Standards} and \ref{fig:Models}.

\section{Spectroscopic Analysis} \label{sec:spec analysis}

Near-infrared spectral analysis is so far the most effective method for classifying late T and early Y dwarfs, as they are very faint in the optical. Transitioning from mid T to late T and early Y spectral types, various features of the near-infrared spectrum change. Notably, the $J$-band peak between 1.2 and 1.4 $\mu$m becomes narrower due to enhanced CH$_4$ absorption, the blue side of the $H$-band peak between 1.4-1.6 $\mu$m becomes steeper to due a combination of greater NH$_3$, CH$_4$, and H$_2$O absorption, and the $K$-band beyond 1.8 $\mu$m becomes less pronounced due to increased H$_2$O absorption.

The near-infrared spectrum of W1055+5443 (Figure \ref{fig:IndivSpectra}) exhibits deep CH$_4$ absorption in the $Y$-, $J$-, $H$-, and $K$-bands, noticeable NH$_3$ absorption in the $H$-band, and strong H$_2$O absorption between the $J$/$H$ and $H$/$K$ bands (see \citealt{Burgasser_2000} for a table of relevant near-infrared absorption features). The W1055+5443 spectrum shows $Y$, $J$, $H$, and $K$ flux peaks  that continuously decrease in peak amplitude. Additionally, W1055+5443 shows unusually high fluxes in the $Y$-, $H$-, and $K$-bands, which do not align with the Y0 standard that fits quite well at $J$-band (see Figure \ref{fig:Standards}). The high W1055+5443 $K$-band amplitude and concavity resemble characteristics seen in mid-T dwarf standards, while the NH$_3$ absorption and steeper slope in the $H$-band are typical of dwarfs with spectral types of T9 or later, as is the narrow $J$-band. 

The $Y$-band flux peak is anomalously high relative to the $J$-band peak, a characteristic not seen in T dwarfs and rarely observed in Y dwarfs. Typically, the $K$-band flux in Y dwarfs is almost entirely suppressed due to enhanced H$_2$ collision-induced absorption (CIA) at higher gravity, but this is not the case for W1055+5443. The W1055+5443 spectrum displays a prominent potential absorption feature centered at 1.015 $\mu$m which is also observed in the near-infrared spectrum of WISEP 182831.08+265037.8, where it was tentatively attributed to CH$_4$ \citep{Cushing_2021}. Due to these peculiarities, we conducted both individual dwarf and binary modeling analyses for W1055+5443 (see Section \ref{subsec:models}).

The unusual nature of W1055+5443's spectrum calls for measurements of multiple spectral indices, comparison against standards, and comparison with models. W1055+5443 would benefit from additional $Y$, $H$, and $K$ broadband photometric detections as crosschecks on the accuracy of the NIRES reduction's relative flux calibration across spectral orders.

\subsection{Spectral Classification\label{subsec:j_class}}

\begin{figure*}[ht]
\includegraphics[scale=0.7]{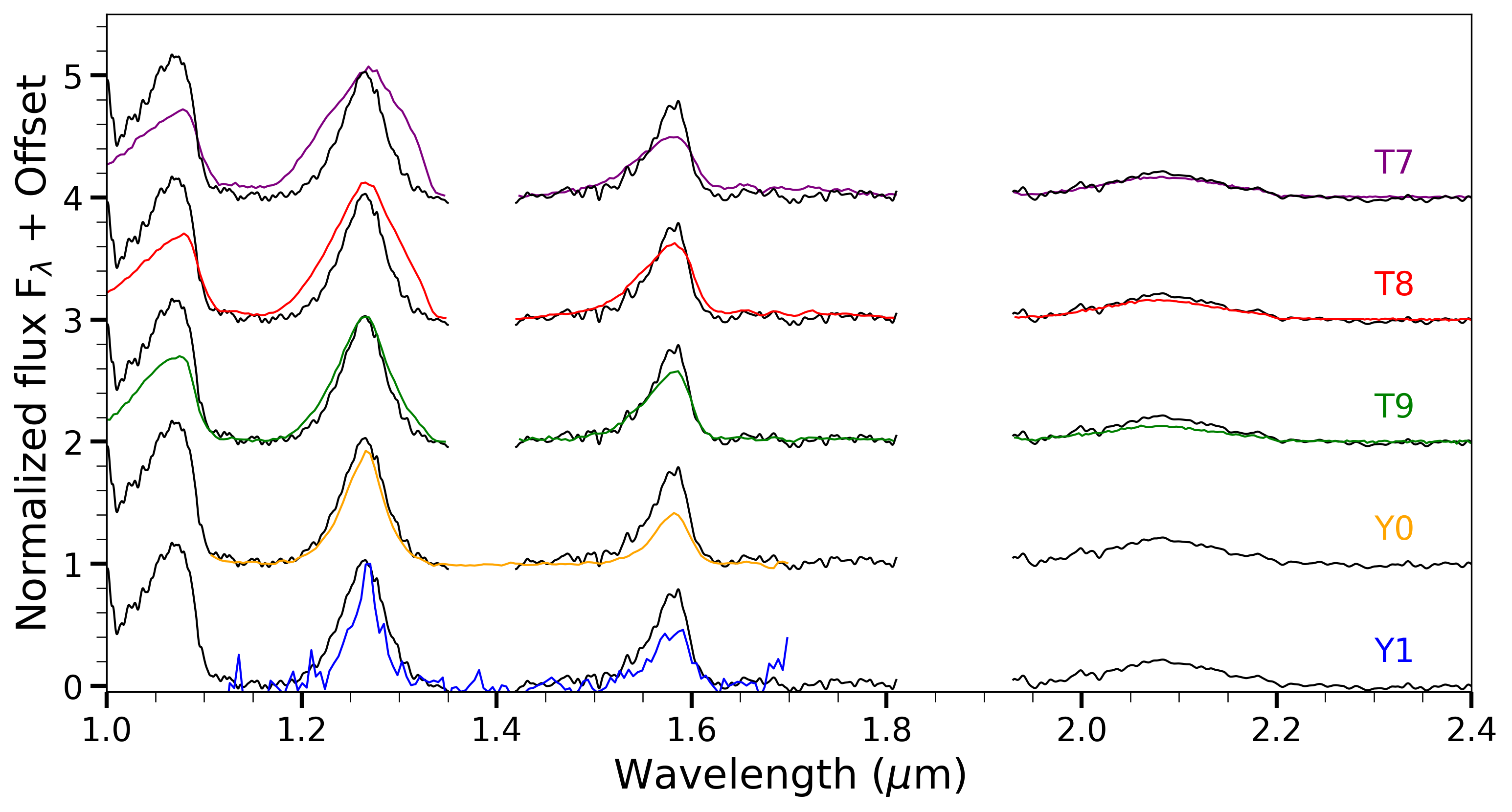}
\caption{\label{fig:Standards} Keck/NIRES near-infrared spectrum of CWISE J105512.11+544328.3 (black) plotted against spectral standards: T7 \citep{Burgasser_2006}, T8 \citep{2004AJ....127.2856B}, T9 \citep{Cushing_2011}, Y0 \citep{Cushing_2011}, and Y1 \citep{Kirkpatrick_2012}.
}
\end{figure*}

We compare the spectrum of W1055+5443 against late T and early Y brown dwarf spectral standards. The comparison is depicted in Figure \ref{fig:Standards}, where W1055+5443 is overplotted along with spectral standards of type T7 and later. We utilized near-infrared spectra for the T7, T8, and T9 standards from the SPLAT SpeX prism library  \citep{burgasser2017spex} as recommended by \cite{Kirkpatrick_2010}, providing data in the $Y$-, $J$-, $H$-, and $K$- bands. The spectrum of W1055+5443 was smoothed to a lower resolution comparable to that of the standards. The Y0 and Y1 standards are drawn from \cite{Cushing_2011} and \cite{Kirkpatrick_2012} respectively, in which the Y0 and Y1 standards have data for the $J$- and $H$- bands.

By focusing on the $J$-band as the most crucial factor for classification, we observed that the $J$-band peaks of T7 and T8 dwarf standards were significantly wider than that of W1055+5443. Whereas the T9 spectral standard's $J$-band peak has a base width slightly broader than our object's, the Y0 dwarf standard closely matches our object's entire $J$-band peak (with only a very slight difference in width). In contrast, the Y1 dwarf standard appears too narrow and slightly offset toward the red in $J$-band. 

As previously mentioned, the $Y$-band of W1055+5443 is unusual, with a higher peak flux than in $J$-band and a strong feature at 1.015~$\mu$m possibly attributable to CH$_4$. Unfortunately, the Y0 and Y1 standards lack $Y$-band data for comparison. A $Y$-band spectral morphology like that of W1055+5443 is sometimes seen in the spectra of other Y0 dwarfs, whereas Y1 dwarfs have steeper slopes at $Y$-band at $\sim$1.05~$\mu$m (see Figures 9 and 11 of \citealt{Schneider_2015}). This  again argues for classification of W1055+5443 as a Y0 dwarf.

The $H$-band of W1055+5443 exhibits peculiar characteristics in terms of its high peak amplitude and its blue side slope. A signature of the T/Y boundary is the enhanced imprint of NH$_3$ absorption on the blue side of the $H$-band peak toward lower temperatures \citep[e.g.,][]{Cushing_2011}. W1055+5443 displays a relatively high blue-side $H$-band slope similar to that expected for a Y0 dwarf. However, the higher $H$-band peak flux of W1055+5443 compared to all of the Figure \ref{fig:Standards} standards prevents us from drawing definitive spectral typing conclusions from the $H$-band.

The $K$-band flux of W1055+5443 is surprisingly high. As temperature decreases in typical brown dwarfs, the $K$-band flux is expected to decrease relative to the $J$-band and flatten due to increased absorption by H$_2$O and CH$_4$, as seen from the T7 and later standards (Figure \ref{fig:Standards}). However, the $K$-band spectrum of W1055+5443 appears better matched to that of a mid-late T dwarf rather than a Y dwarf, yet it is still not a perfect match. While a very low gravity could help explain the elevated $K$-band of W1055+5443 (Section \ref{subsec:models}), objects near the T/Y boundary are generally expected to be old in age, which seemingly would not align with the narrative of a low gravity, and hence young, object. An alternative hypothesis could be that W1055+5443 is a binary, but this notion is deemed implausible upon further investigation in Section \ref{subsec:models}. Based on our comparisons against brown dwarf standards, we assign W1055+5443 a spectral type of Y0 (pec) $\pm$ 0.5.

\subsection{Spectral index measurements\label{subsec:indices}}

Spectral index measurements play a crucial role in cases where spectra are challenging to interpret. Although visual classification methods are generally preferred over index-based classifications (as emphasized in \citealt{Kirkpatrick_2012}), we find that spectral indices provide additional support for a Y0 classification in the case of W1055+5443. Table \ref{tab:indices} presents all of our measured spectral indices for W1055+5443. Our spectrum covers the entire 1-2.4~$\mu$m wavelength range, enabling us to measure more indices than typically available for a Y dwarf (see Figure 7 of \citealt{Cushing_2011} for a visualization of the wavelength ranges contributing to these indices).

We measured several indices involving the $J$-, $H$-, and $K$- bands including NH$_3$-$H$, $K$/$J$ \citep{2008A&A...482..961D}, W$_J$ \citep{2007MNRAS.381.1400W}, H$_2$O-$J$, CH$_4$-$J$, H$_2$O-$H$, CH$_4$-$H$, H$_2$O-$K$, CH$_4$-$K$ \citep{Burgasser_2006}, and the $J$ narrow index \citep{Mace_2013}. All of the indices support a classification later than T7, with a few specifically suggesting a classification of Y0. Notably, the CH$_4$-$J$ index, which measures the decline of the red side of the $J$-band, strongly favors a Y0 dwarf classification, aligning with our visual inspection and matching the spectral index figures presented in \cite{Cushing_2011}.

\begin{table}[htbp]
\raggedright
\caption{Spectral Indices of CWISE J105512.11+544328.3}
\label{tab:indices}
\begin{tabular}{lll}
\hline\hline
Index & Value$^a$ & Corresponding \\
       &       & Spectral Type$^b$ \\
\hline

NH$_3$-$H$ (1) & 0.427  $\pm$ 0.0012 & Y0 \\
NH$_3$-$H$ (2) & 0.741  $\pm$ 0.0012 & Y0 \\
CH$_4$-$J$& 0.0385 $\pm$ 0.0007 & Y0 \\
H$_2$O-$J$& 0.0168 $\pm$ 0.0029 & $>$ T8.5 \\
CH$_4$-$H$& 0.0509 $\pm$ 0.0007 & $>$ T8.5 \\
H$_2$O-$H$& 0.0944 $\pm$ 0.0019 & $>$ T8 \\
$J$ narrow & 0.820 $\pm$ 0.0389 & $>$ T8\\
W$_J$      & 0.298  $\pm$ 0.0030 & $>$ T7 \\
$K$/$J$      & 0.234  $\pm$ 0.0003 &  \\
H$_2$O-$K$ & 0.289  $\pm$ 0.0010 &  \\
CH$_4$-$K$ & 0.0312 $\pm$ 0.0004 &  \\
\hline
\end{tabular} \\[1ex]
\footnotesize \textbf{$^a$} Values calculated using a Gaussian smoothing of $\sigma$ = 5 pixels. \\
\footnotesize \textbf{$^b$} Corresponding spectral types from \cite{Cushing_2011} and \cite{Kirkpatrick_2012}. \\
\footnotesize (1) Original NH$_3$-$H$ index from \cite{2008A&A...482..961D}. \\
(2) New NH$_3$-$H$ index proposed in \cite{2021A&A...655L...3M}.
\end{table}

\begin{figure*}[ht]
\includegraphics[scale=0.59]{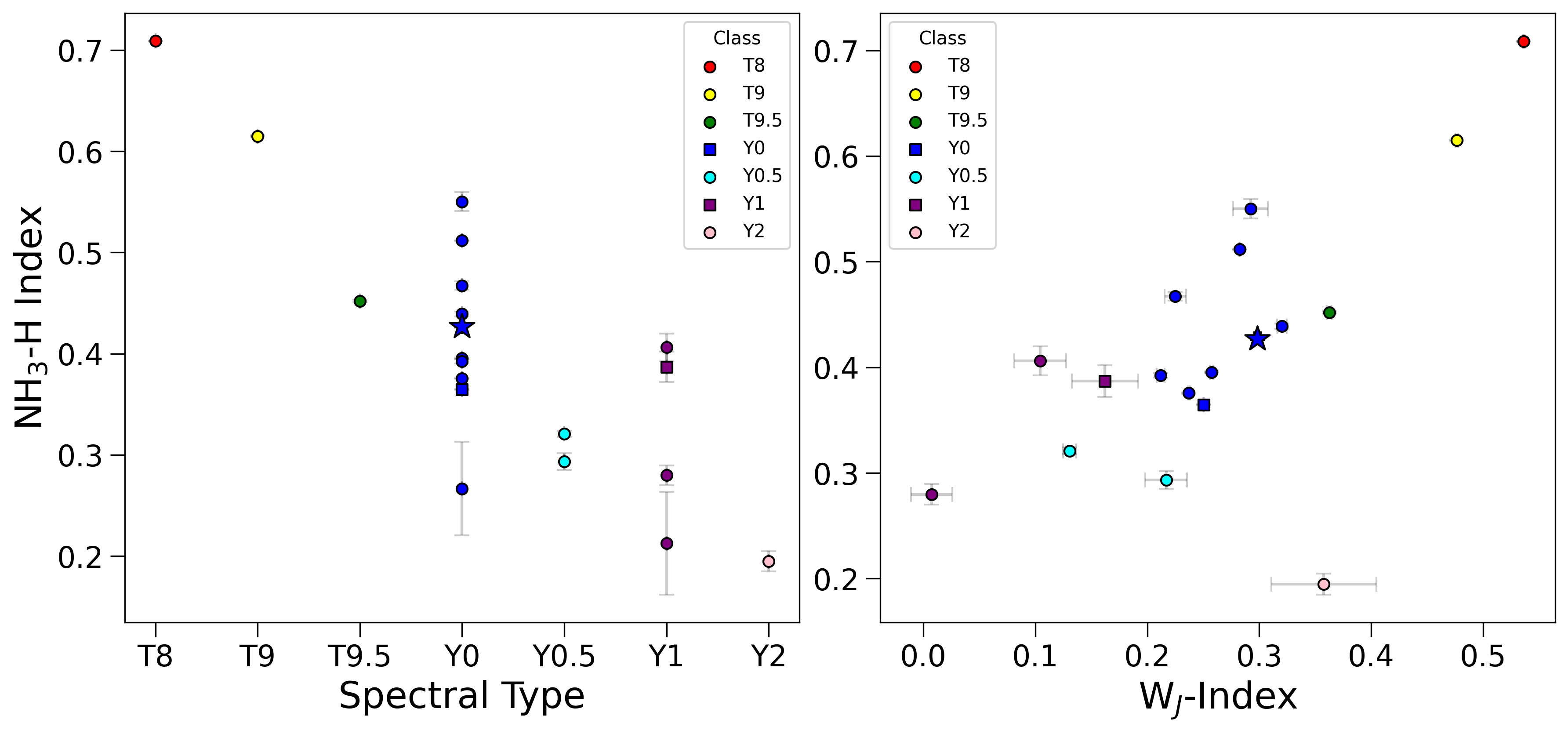}
\caption{\label{fig:NH3_WJ} Eighteen T8 to Y2 dwarfs from \cite{Schneider_2015} plotted with  W1055+5443 (blue star). Left panel: \cite{2008A&A...482..961D} NH$_3$-$H$ index versus spectral type. Right panel: \cite{2008A&A...482..961D} NH$_3$-$H$ index plotted against W$_J$ index. In both panels, spectral type defines each data point's color and square symbols are spectral standards.}
\end{figure*}

The NH$_3$-$H$ index is arguably the most critical index for classifying late T and early Y dwarfs. As temperature decreases below 600~K in brown dwarfs, ammonia absorption becomes more dominant and detectable in the blue side of the $H$-band peak. At temperatures below 350~K, atmospheric ammonia is predicted to begin condensing. Our NH$_3$-$H$ index measurement of 0.427 $\pm$ 0.0012 for W1055+5443 aligns with expectations for an early Y dwarf. To visualize the NH$_3$-$H$ index as a function of spectral class near the T/Y boundary, we plotted the NH$_3$-$H$ index of eighteen T8-Y2 dwarfs from \cite{Schneider_2015} in the left panel of Figure \ref{fig:NH3_WJ}, along with our NH$_3$-$H$ index calculated for W1055+5443 (see Figure 2 of \citealt{Cushing_2021} for an expanded spectral type range). Although there is a fair amount of scatter for the Y0 and Y1 dwarfs, a clear slope is seen from T8 to Y2 and the position of W1055+5443 most closely corresponds to a type of Y0.

The best method of measuring NH$_3$ in near-infrared brown dwarf spectra remains a subject of ongoing work, with some measurements of NH$_3$ being performed at $\approx$ 1.5-1.6~$\mu$m. A new approach focusing on the 1.5-1.61 $\mu$m wavelength range, derived in \cite{2021A&A...655L...3M} based on simulated spectra, offers a potentially improved mapping from spectral index to spectral type (see Figure 2 of \citealt{2021A&A...655L...3M}). Using this method, we obtain NH$_3$-$H$ = 0.741 $\pm$ 0.0012 (\citealt{2021A&A...655L...3M} convention), closely aligning with spectral type Y0.

Another valuable index for T/Y classification is the W$_J$ index, which probes the ammonia and methane absorption between 1.18-1.285 $\mu$m. By itself, the W$_J$ index does not offer major insights into the spectral type of W1055+5443, suggesting only a spectral type later than T7. However, when compared with the NH$_3$-$H$ index as discussed in \cite{2008A&A...482..961D}, we observe a strong correlation between W$_J$ and NH$_3$-$H$. To illustrate this correlation, we plotted NH$_3$-$H$ against W$_J$ for W1055+5443 and the eighteen aforementioned T8-Y2 dwarfs (right panel of Figure \ref{fig:NH3_WJ}). While there is significant scatter in the location of Y0 dwarfs, our object's position is reasonably centered within the data points belonging to other Y0 dwarfs, further supporting its classification as Y0.

As for other spectral indices such as H$_2$O-$J$, CH$_4$-$H$, H$_2$O-$H$, $J$ narrow, $K$/$J$, H$_2$O-$K$, and CH$_4$-$K$, they either lack sufficient prior literature measurements for objects near the T/Y boundary or their trends do not provide a clear indication of spectral type beyond the mid-late T regime. As  measurements and spectra of brown dwarfs improve in the future, these indices may become more valuable in the evaluation of spectral types and properties of extremely cold brown dwarfs.

\subsection{Model Comparisons}\label{subsec:models}

To understand the physical properties of W1055+5443, such as effective temperature, gravity, and metallicity, we compared its near-infrared spectrum to various LOWZ models \citep{Meisner_2021} and Sonora Bobcat models \citep{2021ApJ...920...85M}, both of which incorporate low temperatures and low metallicities. The original LOWZ models cover a grid with parameters sampled as follows: $-$2.5 $\leq$ [m/H] $\leq$ +1.0 in steps of 0.25 and 0.5, 500~K $\leq$ $T_{\rm eff}$ $\leq$ 1600~K in steps of 50~K and 100~K, 3.5 $\leq$ log(g) $\leq$ 5.25 (with g in cgs) in steps of 0.25 and 0.5, three C/O values with 0.1 $\leq$ C/O $\leq$ 0.85, and three $K_{zz}$ values with $-$1 $\leq$ log($K_{zz}$) $\leq$ 10. We also incorporated an extension of the LOWZ models (M. Line, priv. comm.), which continues the original grid to lower temperatures of 400~K, 375~K, 350~K, 325~K, and 300~K. To identify the best-fit model of W1055+5443, we compared its near-infrared spectrum to all 8,582 LOWZ models using the standard $\chi^2$ metric. During fitting, each model spectrum was initially normalized to unity at its $J$-band peak, and subsequently the overall normalization of each model at its $J$-band peak was treated as a free parameter between 0.3 and 3.0, following \cite{Meisner_2021}.

\begin{figure*}[ht]
\includegraphics[scale=0.77]{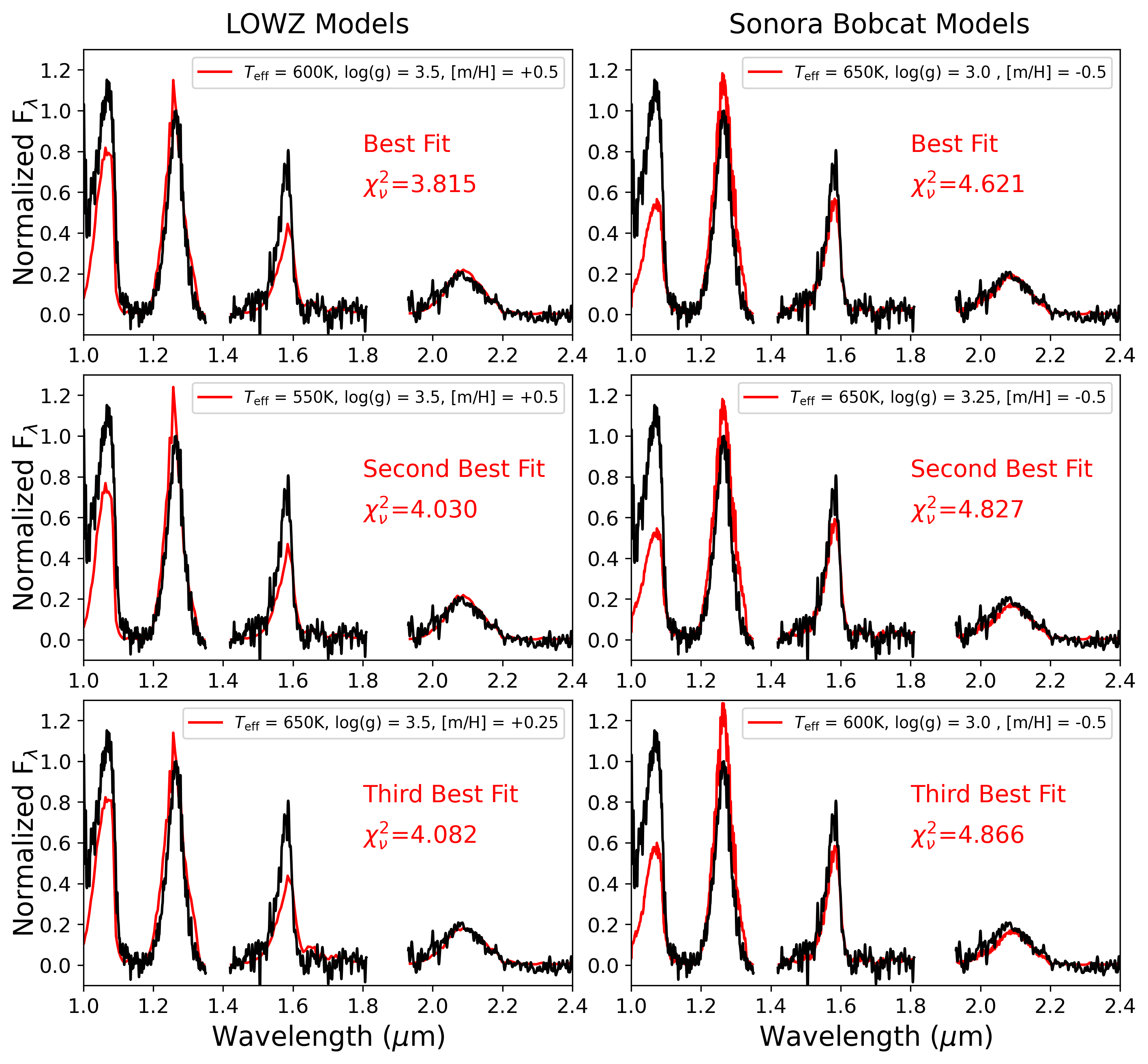}
\caption{\label{fig:Models} Comparison of the W1055+5443 near-infrared spectrum normalized between 1.27 and 1.29 $\mu$m (black) versus various fitted models (red). The left models are the top three best-fitting LOWZ models from \cite{Meisner_2021}. The right models are the top three best-fitting Sonora Bobcat models from \cite{2021ApJ...920...85M}. Both of these model grids explore multiple values of $T_{\rm eff}$, log(g), and [m/H]. The LOWZ models shown have log(K$_{\rm zz}$) = 2.0, while the Sonora Bobcat models are not parameterized in terms of K$_{\rm zz}$.}
\end{figure*}

The left column of Figure \ref{fig:Models} shows our three best-fitting LOWZ models for W1055+5443. All three best-fitting LOWZ models have an eddy diffusion log(K$_{zz}$) = 2.0 and subsolar C/O = 0.1. All three best-fit LOWZ models have the lowest available log(g) = 3.5, which does not match expectations for an anticipated old-age Y dwarf. Two of the best-fit models prefer a supersolar metallicity ([m/H] = +0.5), while the third one indicates a slightly supersolar metallicity ([m/H] = +0.25), in contrast to the subsolar C/O ratio.

The $T_{\rm eff}$ values of 600~K, 550~K, and 650~K for the three best-fitting LOWZ models are higher than expected for a Y dwarf. While the LOWZ models fit relatively well in the $J$- and $K$-bands, they consistently fail to match the $Y$- and $H$-bands. To investigate these discrepancies, we performed model comparisons with varying temperatures restricted to individual bands, revealing that the $Y$- and $J$-bands are better matched with lower temperature models (400~K $\leq$ $T_{\rm eff}$ $\leq$ 550~K), while the $H$- and $K$-bands align better with higher temperature models (700~K $\leq$ $T_{\rm eff}$ $\leq$ 850~K). When fitting for LOWZ models that best match at the $J$-band, the most important band in our visual classification process (Section \ref{subsec:j_class}), we find that temperatures between 500~K and 600~K fit best. No single LOWZ model offers a satisfactory fit to all bands.

To potentially find a better-fitting model, we also pursued comparisons against the Sonora Bobcat grid \citep{2021ApJ...920...85M}. These models have the following parameters: 3.0 $\leq$ log(g) $\leq$ 5.5 (with g in cgs) and 200~K $\leq$ $T_{\rm eff}$ $\leq$ 2400~K. Steps in $T_{\rm eff}$ vary from 25~K to 100~K and steps in log(g) are 0.25 or 0.5. Models are provided for [m/H] = $-$0.5, 0.0, and +0.5. We chose Sonora Bobcat models because they offer a broader range of gravities and more $T_{\rm eff}$ gradations between 300-600~K, the temperature range we expect for W1055+5443 given its near-infrared spectral classification. Like the LOWZ models, we compared the Sonora Bobcat spectra to W1055+5443 using a standard $\chi^2$ metric and a free overall multiplicative normalization parameter for each model. The Sonora Bobcat fitting results are presented in the right-hand column of Figure \ref{fig:Models}. We note that the Sonora Bobcat and LOWZ models may differ in methane line lists, possibly resulting in different metallicity and model results. Both model grids are cloudless and under chemical equilibrium.

The best three Sonora Bobcat fits indicate $T_{\rm eff}$ = 650~K and 600~K, log(g) = 3.0 and 3.25, and [m/H] = $-$0.5. The temperatures are higher than we anticipate for a Y dwarf, but this can be attributed to the $H$- and $K$-bands, which drag the preferred $T_{\rm eff}$ higher relative to the $Y$- and $J$-bands. As with the LOWZ models, none of the Sonora Bobcat models fit all the spectral features of W1055+5443 well. The best-fitting Sonora Bobcat models show better fits in the $J$- and $H$-bands while the best-fitting LOWZ models provide superior fits for the $Y$- and $K$-bands. Some of these differences may arise from the fact that the LOWZ models do not offer gravities below 3.5, while the Sonora Bobcat models extend to log(g) values of 3.0. 

Interestingly, for both model grids employed, the lowest gravity models consistently fit W1055+5443 the best. This is quite unexpected for an extremely cold brown dwarf, which we would typically assume to be older in age. If even lower-gravity models were available, they could perhaps result in yet better fits to the W1055+5443 spectrum. Additionally, we checked the \cite{2023ApJ...950....8L} cloudy and cloudless equilibrium and disequilibrium chemistry models intended for Y dwarfs using our same $\chi^2$ procedure and found no significantly improved fit. Based on our model analysis, we tentatively assign W1055+5443 a log(g) $\leq$ 4.5. However, our  results indicate that no single model (among those we utilized) can perfectly fit all the spectral features of W1055+5443; further investigation is needed to better elucidate this anomalous object's physical properties.

\subsubsection{Binary Modeling}\label{Binary}

In instances where the relatively short wavelength portion of a brown dwarf system's spectrum seems to favor a different type/temperature than its relatively long wavelength portion, it is natural to suspect binarity \citep[e.g.,][]{Daniella_binaries}. We therefore investigated whether it is possible to better explain the W1055+5443 spectrum as a superposition of two model spectra rather than as a single brown dwarf. We limited our binarity analysis to component models below 700~K, as individual dwarf models above this temperature are too wide in the $J$-band to result in any high-quality binary model of W1055+5443.

We performed a comparison against 5460 Sonora Bobcat model combinations with $T_{\rm eff}$ $<$ 700~K and varying log(g). Since [m/H] = $-$0.5 provided the best fit for individual Sonora Bobcat models (and due to computational constraints), we kept the metallicity fixed at [m/H] = $-$0.5. None of the model combinations matched the spectrum of W1055+5443 very well. As a result, we conclude that W1055+5443 is not part of a brown dwarf binary system with component properties in the parameter space that we tested (300 $\leq$ $T_{\rm eff}$ $<$ 700~K, 3.0 $\leq$ log(g) $\leq$ 5.25, [m/H]= $-$0.5). An L dwarf companion is implausible because it would require the system to have, for instance, a W2 apparent mag at least $\sim 3+$ magnitudes brighter than we observe given the W1055+5443 distance of $\sim 7$ pc \citep{Kirkpatrick_2021}. Even a typical T8 dwarf ($T_{\rm eff} \approx 700$~K) at the trigonometric distance of W1055+5443 would be expected to yield an apparent W2 magnitude $> 1$ mag brighter than observed \citep{Kirkpatrick_2021}.

\section{Discussion} \label{sec:discussion}

\subsection{$T_{\rm eff}$ and Spectral Type Polynomials}\label{subsec:properties}

Given our imperfect model fitting results for W1055+5443, we considered additional options for estimating quantities such as its effective temperature and spectral type, as a matter of due diligence. We employ the polynomial relations from \cite{Kirkpatrick_2019, Kirkpatrick_2021}, which allow us to obtain spectral type and temperature estimates from absolute magnitudes and colors.

We performed polynomial spectral type estimates for W1055+5443 using the following inputs: $M_{\rm ch2}$, $M_{\rm ch1}$, $M_J$, $J$$-$ch2, and ch1$-$ch2. We find the following results: $M_{\rm ch2}$: Y0.4 $\pm$ 1.3, $M_{\rm ch1}$: T9.8 $\pm$ 0.9, $M_J$: T9.6 $\pm$ 0.5, $J$$-$ch2: T9.1 $\pm$ 0.5, and ch1$-$ch2: T8.0 $\pm$ 1.3. These results are reasonably consistent with our Y0 (pec) $\pm$ 0.5 spectral classification (Section \ref{subsec:j_class}). The photometric type estimate based on ch1$-$ch2 color is notably discrepant with our spectroscopic classification, which is not surprising given W1055+5443's outlier status in Figure \ref{fig:ch12Sp}.

Similarly, polynomial relations allow us to estimate $T_{\rm eff}$ from spectral type, $M_{W2}$, $M_{\rm ch2}$, and ch1$-$ch2. Using Y0 $\pm$ 0.5 as the spectral type input, we would expect $T_{\rm eff}$ = 420.2 $\pm$ 142.7~K. The central $T_{\rm eff}$ estimates from $M_{W2}$ and $M_{\rm ch2}$ are slightly lower, yielding $T_{\rm eff}$ = 395.9 $\pm$ 73.4~K and $T_{\rm eff}$ = 393.0 $\pm$ 76.6~K, respectively. Finally, we find that the object's anomalously blue ch1$-$ch2 color yields an estimate of $T_{\rm eff}$ = 660.2 $\pm$ 83.2~K, which aligns better with the upper end of our best-fit temperatures derived via model fits. Combining the modeling results and polynomial calculations, we assign a rather uncertain $T_{\rm eff}$ = 500 $\pm$ 150~K for W1055+5443.

\subsection{Kinematics and age}\label{subsec:kin}

We used an updated version of the BANYAN~$\Sigma$ tool \citep{Gagné_2018} to determine whether W1055+5443 may be a member of a known nearby young association. The BANYAN~$\Sigma$ tool includes spatial-kinematic models for several nearby young associations (originally 27 such groups), and allows users to determine whether a star is a likely member of these groups based on the observed kinematic measurements, with a Bayesian model selection method. The tool allows to determine membership probabilities with missing parallax or radial velocity, and in such cases the marginalization integrals are solved analytically. In our use case, W1055+5443 benefits from a parallax (see Table~\ref{tab:properties}), therefore only the marginalization integral on the radial velocity was involved. We used an updated set of models in BANYAN~$\Sigma$ (Gagn\'e et al., in preparation), which includes more recently discovered associations such as the $\mu$~Tau association \citep{2020ApJ...903...96G} and the Crius associations of \cite{2022ApJ...939...94M}, as well as open clusters at distances up to 500\,pc from the Sun.

We find that W1055+5443 obtains a 98.2\%\ membership probability in Crius~197, with an optimal distance of $6.60 \pm 0.07$\,pc, and an optimal radial velocity of $7.2 \pm 0.7\,\kms$. These quantities represent the values that would optimize the membership probability (not including the parallax measurement for the former), but measured values slightly outside of these range do not guarantee non-membership because the Crius~197 model spans 0.8\,\kms\ in $UVW$ space and 19\,pc in $XYZ$ space. We find that W1055+5443 falls at a distance of 23.2\,pc from the center of the $XYZ$ projection of the BANYAN~$\Sigma$ model, and at 1.9\,\kms\ from the center of its $UVW$ projection. These numbers are indicative of a very good kinematic match to the model, but do not indicate that 98.2\%\ of such candidates would be real members, because the BANYAN~$\Sigma$ Bayesian probabilities are normalized to obtain a recovery rate (true positives) of 90\%\ with a selection cut of $> 90$\%\ membership probabilities for convenience with all-sky searches. Obtaining a radial velocity measurement for W1055+5443 would be useful to corroborate its possible membership, but even if the radial velocity matches that of Crius~197, the contamination rate of the group itself and its physical nature still need to be established with care.

Crius~197 is a nearby candidate moving group discovered by \cite{2022ApJ...939...94M} with the HDBSCAN clustering method \citep{campello2013density} applied on a nearby sample of Gaia~DR2 \citep{2018A&A...616A...1G} stars with full kinematics. The initial group contained 10 stars, but was not part of the sample that was studied in more detail by \cite{2022ApJ...939...94M} because the median absolute deviation of its membership distribution across the Galactic coordinate $Z$ was measured at 19.9\,pc, above the 15\,pc threshold based on typical nearby associations.

A preliminary analysis based on the TESS \citep{TESS} rotation periods of the Crius~197 members yields an estimated age of $180 \pm 9$\,Myr, mostly constrained by the 5 members LP~276--29~A, PM~J11214--2645, PM~J11591--7616, LP~845--14, and PM~J08355--2200 (Moranta et al., in preparation), but the group also contains two members that appear to be outliers in the rotation period--color sequence (HIP~42333, $P_{\rm rot} \approx 8$\,days, i.e. $\approx 670$\,Myr; HD~80622, $P_{\rm rot} \approx 12$\,days, i.e. $\gtrapprox 1$\,Gyr). It is still unclear whether those stars are simply chance interlopers to the association, or whether they indicate that Crius~197 is a heterogeneous population.

If we assume that W1055+5443 has an age of $\approx$ 180\,Myr with an estimated $\log\left(L_{\rm bol}/L_{\rm Sun}\right) = -6.0 \pm 0.1$ (based on the method of \citealt{2015ApJ...810..158F} applied to the data in Table~\ref{tab:properties}), we estimate that it would have a mass of about 4--6\,\mjup, well within the planetary-mass regime. However, we consider this a tentative assessment that requires future confirmation.

\subsection{Conclusion}

We have presented photometric and spectroscopic analyses of CWISE J105512.11+544328.3 incorporating archival survey photometry, literature photometry, literature astrometry, and new follow-up near-infrared spectroscopy from Keck/NIRES. We find that W1055+5443 best matches the Y0 spectral standard, particularly at $J$-band, but does not match well with any brown dwarf standard across the full 1-2.4~$\mu$m wavelength range available from NIRES. Oddly, the $K$-band of W1055+5443 aligns well with that of mid-late T dwarfs; binarity is unlikely as a potential explanation for the later spectral types favored at $YJ$ compared to the earlier spectral types favored at $HK$. Our W1055+5443 spectrum shows strong ammonia absorption plus methane and water absorption in the $H$-band consistent with a Y-type classification, though the W1055+5443 $H$-band peak amplitude is anomalously high. NH$_3$-$H$ and CH$_4$-$J$ spectral indices point to a Y0 classification, while other indices are all consistent with a type later than T7. We assign W1055+5443 a spectral type of Y0 (pec) $\pm$ 0.5.

Our investigation of W1055+5443's physical properties using LOWZ and Sonora Bobcat atmospheric models finds that no such model fits the entirety of W1055+5443's near-infrared spectrum well. The best-fitting Sonora Bobcat models all have subsolar metallicities of [m/H] = $-$0.5 and very low gravities of 3.0 $\leq$ log(g) $\leq$ 3.25. These Sonora Bobcat models fit the $J$-, $H$-, and $K$-bands reasonably well, but not the $Y$-band. The best-fitting LOWZ models favor supersolar metallicities of [m/H] = +0.25 or +0.5 and low gravities of log(g) = 3.5, matching the $Y$-band data better than the Sonora Bobcat models though still far from perfectly. We are unable to effectively assign W1055+5443 a metallicity, but we tentatively assign a log(g) $\leq$ 4.5. Considering a combination of spectroscopic model fits and photometric estimates, we assign W1055+5443 an uncertain temperature of $T_{\rm eff}$ = 500 $\pm$ 150~K.

W1055+5443 has the bluest Spitzer ch1$-$ch2 color of any spectroscopically confirmed Y dwarf, and is also a modest photometric outlier along a number of axes (see Figures \ref{fig:W23vW12}-\ref{fig:ch12J}). Recent studies have suggested that the combination of relatively blue Spitzer ch1$-$ch2 color and extremely cold temperature may be characteristic of low metallicity (see especially Figure 7 of \citealt{Meisner_2021} and Figure 2 of \citealt{The_Accident}). However, spectroscopic and kinematic considerations do not particularly favor low metallicity as the explanation for W1055+5443's anomalous properties. Although the Sonora Bobcat models suggest low metallicity for W1055+5443, LOWZ models prefer the opposite. The pronounced $K$-band concavity of our W1055+5443 spectrum stands in contrast to the flattened $K$-band that would be expected for an extremely cold and low-metallicity dwarf \citep[e.g.,][]{Zhang_2019} and the very low gravities indicated by both sets of models employed are opposite of high-gravity expectations for an old, low-metallicity dwarf. The W1055+5443 tangential velocity of $\approx 50$~km~s$^{-1}$ is high relative to the solar neighborhood median value for brown dwarfs \citep{Kirkpatrick_2021}, but not large enough to strongly indicate thick disk or halo membership \citep{Faherty_BDKP}.

\section{Acknowledgments}

We thank Mike Line for providing a low-temperature extension of the LOWZ model grid and we thank the anonymous referee for helpful comments which have improved this manuscript.

 \ \ This work has been supported in part by the NASA Citizen Science Seed Funding Program, Grant 80NSSC21K1485. This material is based upon work supported by the National Science Foundation under Grant No. 2007068, 2009136, and 2009177.

The work of Grady Robbins and Aaron Meisner is supported by NOIRLab, which is managed by the Association of Universities for Research in Astronomy (AURA) under a cooperative agreement with the National Science Foundation.

Some of the data presented herein were obtained at the W. M. Keck Observatory, which is operated as a scientific partnership among the California Institute of Technology, the University of California, and the National Aeronautics and Space Administration. The Observatory was made possible by the generous financial support of the W. M. Keck Foundation. The authors wish to recognize and acknowledge the very significant cultural role and reverence that the summit of Maunakea has always had within the indigenous Hawaiian community.  We are most fortunate to have the opportunity to conduct observations from this mountain.

This publication makes use of data products from the Wide-field Infrared Survey Explorer, which is a joint project of the University of California, Los Angeles, and the Jet Propulsion Laboratory/California Institute of Technology, funded by the National Aeronautics and Space Administration. This publication also makes use of data products from NEOWISE, which is a project of the Jet Propulsion Laboratory/California Institute of Technology, funded by the Planetary Science Division of the National Aeronautics and Space Administration.

\vspace{5mm}
\facilities{WISE,
Keck/NIRES
}

\software{
SPLAT \citep{burgasser2017spex}, 
WiseView \citep{2018ascl.soft06004C},
WSA \citep{wsa}
}

\bibliography{AcademicBib}

\begin{thebibliography}{}
\expandafter\ifx\csname natexlab\endcsname\relax\def\natexlab#1{#1}\fi
\providecommand{\url}[1]{\href{#1}{#1}}
\providecommand{\dodoi}[1]{doi:~\href{http://doi.org/#1}{\nolinkurl{#1}}}
\providecommand{\doeprint}[1]{\href{http://ascl.net/#1}{\nolinkurl{http://ascl.net/#1}}}
\providecommand{\doarXiv}[1]{\href{https://arxiv.org/abs/#1}{\nolinkurl{https://arxiv.org/abs/#1}}}

\bibitem[{{Bardalez Gagliuffi} {et~al.}(2014){Bardalez Gagliuffi}, {Burgasser}, {Gelino}, {Looper}, {Nicholls}, {Schmidt}, {Cruz}, {West}, {Gizis}, \& {Metchev}}]{Daniella_binaries}
{Bardalez Gagliuffi}, D.~C., {Burgasser}, A.~J., {Gelino}, C.~R., {et~al.} 2014, \apj, 794, 143, \dodoi{10.1088/0004-637X/794/2/143}

\bibitem[{{Beichman} {et~al.}(2013){Beichman}, {Gelino}, {Kirkpatrick}, {Barman}, {Marsh}, {Cushing}, \& {Wright}}]{Beichman_2013}
{Beichman}, C., {Gelino}, C.~R., {Kirkpatrick}, J.~D., {et~al.} 2013, \apj, 764, 101, \dodoi{10.1088/0004-637X/764/1/101}

\bibitem[{Burgasser {et~al.}(2006)Burgasser, Geballe, Leggett, Kirkpatrick, \& Golimowski}]{Burgasser_2006}
Burgasser, A.~J., Geballe, T.~R., Leggett, S.~K., Kirkpatrick, J.~D., \& Golimowski, D.~A. 2006, The Astrophysical Journal, 637, 1067, \dodoi{10.1086/498563}

\bibitem[{{Burgasser} {et~al.}(2004){Burgasser}, {McElwain}, {Kirkpatrick}, {Cruz}, {Tinney}, \& {Reid}}]{2004AJ....127.2856B}
{Burgasser}, A.~J., {McElwain}, M.~W., {Kirkpatrick}, J.~D., {et~al.} 2004, \aj, 127, 2856, \dodoi{10.1086/383549}

\bibitem[{Burgasser \& the SPLAT Development~Team(2017)}]{burgasser2017spex}
Burgasser, A.~J., \& the SPLAT Development~Team. 2017.
\newblock \doarXiv{1707.00062}

\bibitem[{Burgasser {et~al.}(2000)Burgasser, Wilson, Kirkpatrick, Skrutskie, Colonno, Enos, Smith, Henderson, Gizis, Brown, \& Houck}]{Burgasser_2000}
Burgasser, A.~J., Wilson, J.~C., Kirkpatrick, J.~D., {et~al.} 2000, The Astronomical Journal, 120, 1100, \dodoi{10.1086/301475}

\bibitem[{Campello {et~al.}(2013)Campello, Moulavi, \& Sander}]{campello2013density}
Campello, R.~J., Moulavi, D., \& Sander, J. 2013, in Pacific-Asia conference on knowledge discovery and data mining, Springer, 160--172

\bibitem[{{Caselden} {et~al.}(2018){Caselden}, {Westin}, {Meisner}, {Kuchner}, \& {Colin}}]{2018ascl.soft06004C}
{Caselden}, D., {Westin}, Paul, I., {Meisner}, A., {Kuchner}, M., \& {Colin}, G. 2018, {WiseView: Visualizing motion and variability of faint WISE sources}, Astrophysics Source Code Library, record ascl:1806.004.
\newblock \doeprint{1806.004}

\bibitem[{{Chambers} {et~al.}(2016){Chambers}, {Magnier}, {Metcalfe}, {Flewelling}, {Huber}, {Waters}, {Denneau}, {Draper}, {Farrow}, {Finkbeiner}, {Holmberg}, {Koppenhoefer}, {Price}, {Rest}, {Saglia}, {Schlafly}, {Smartt}, {Sweeney}, {Wainscoat}, {Burgett}, {Chastel}, {Grav}, {Heasley}, {Hodapp}, {Jedicke}, {Kaiser}, {Kudritzki}, {Luppino}, {Lupton}, {Monet}, {Morgan}, {Onaka}, {Shiao}, {Stubbs}, {Tonry}, {White}, {Ba{\~n}ados}, {Bell}, {Bender}, {Bernard}, {Boegner}, {Boffi}, {Botticella}, {Calamida}, {Casertano}, {Chen}, {Chen}, {Cole}, {Deacon}, {Frenk}, {Fitzsimmons}, {Gezari}, {Gibbs}, {Goessl}, {Goggia}, {Gourgue}, {Goldman}, {Grant}, {Grebel}, {Hambly}, {Hasinger}, {Heavens}, {Heckman}, {Henderson}, {Henning}, {Holman}, {Hopp}, {Ip}, {Isani}, {Jackson}, {Keyes}, {Koekemoer}, {Kotak}, {Le}, {Liska}, {Long}, {Lucey}, {Liu}, {Martin}, {Masci}, {McLean}, {Mindel}, {Misra}, {Morganson}, {Murphy}, {Obaika}, {Narayan}, {Nieto-Santisteban}, {Norberg}, {Peacock}, {Pier}, {Postman}, {Primak}, {Rae}, {Rai},
  {Riess}, {Riffeser}, {Rix}, {R{\"o}ser}, {Russel}, {Rutz}, {Schilbach}, {Schultz}, {Scolnic}, {Strolger}, {Szalay}, {Seitz}, {Small}, {Smith}, {Soderblom}, {Taylor}, {Thomson}, {Taylor}, {Thakar}, {Thiel}, {Thilker}, {Unger}, {Urata}, {Valenti}, {Wagner}, {Walder}, {Walter}, {Watters}, {Werner}, {Wood-Vasey}, \& {Wyse}}]{PS1}
{Chambers}, K.~C., {Magnier}, E.~A., {Metcalfe}, N., {et~al.} 2016, arXiv e-prints, arXiv:1612.05560, \dodoi{10.48550/arXiv.1612.05560}

\bibitem[{Cushing {et~al.}(2004)Cushing, Vacca, \& Rayner}]{Cushing_2004}
Cushing, M.~C., Vacca, W.~D., \& Rayner, J.~T. 2004, Publications of the Astronomical Society of the Pacific, 116, 362, \dodoi{10.1086/382907}

\bibitem[{Cushing {et~al.}(2011)Cushing, Kirkpatrick, Gelino, Griffith, Skrutskie, Mainzer, Marsh, Beichman, Burgasser, Prato, Simcoe, Marley, Saumon, Freedman, Eisenhardt, \& Wright}]{Cushing_2011}
Cushing, M.~C., Kirkpatrick, J.~D., Gelino, C.~R., {et~al.} 2011, The Astrophysical Journal, 743, 50, \dodoi{10.1088/0004-637X/743/1/50}

\bibitem[{Cushing {et~al.}(2021)Cushing, Schneider, Kirkpatrick, Morley, Marley, Gelino, Mace, Wright, Eisenhardt, Skrutskie, \& Marsh}]{Cushing_2021}
Cushing, M.~C., Schneider, A.~C., Kirkpatrick, J.~D., {et~al.} 2021, The Astrophysical Journal, 920, 20, \dodoi{10.3847/1538-4357/ac12cb}

\bibitem[{{Cutri} {et~al.}(2013){Cutri}, {Wright}, {Conrow}, {Fowler}, {Eisenhardt}, {Grillmair}, {Kirkpatrick}, {Masci}, {McCallon}, {Wheelock}, {Fajardo-Acosta}, {Yan}, {Benford}, {Harbut}, {Jarrett}, {Lake}, {Leisawitz}, {Ressler}, {Stanford}, {Tsai}, {Liu}, {Helou}, {Mainzer}, {Gettings}, {Gonzalez}, {Hoffman}, {Marsh}, {Padgett}, {Skrutskie}, {Beck}, {Papin}, \& {Wittman}}]{2013wise.rept....1C}
{Cutri}, R.~M., {Wright}, E.~L., {Conrow}, T., {et~al.} 2013, {Explanatory Supplement to the AllWISE Data Release Products}, Explanatory Supplement to the AllWISE Data Release Products, by R. M. Cutri et al.

\bibitem[{{Delorme} {et~al.}(2008){Delorme}, {Delfosse}, {Albert}, {Artigau}, {Forveille}, {Reyl{\'e}}, {Allard}, {Homeier}, {Robin}, {Willott}, {Liu}, \& {Dupuy}}]{2008A&A...482..961D}
{Delorme}, P., {Delfosse}, X., {Albert}, L., {et~al.} 2008, \aap, 482, 961, \dodoi{10.1051/0004-6361:20079317}

\bibitem[{Dye {et~al.}(2017)Dye, Lawrence, Read, Fan, Kerr, Varricatt, Furnell, Edge, Irwin, Hambly, Lucas, Almaini, Chambers, Green, Hewett, Liu, McGreer, Best, Zhang, Sutorius, Froebrich, Magnier, Hasinger, Lederer, Bold, \& Tedds}]{Dye_2017}
Dye, S., Lawrence, A., Read, M.~A., {et~al.} 2017, Monthly Notices of the Royal Astronomical Society, 473, 5113, \dodoi{10.1093/mnras/stx2622}

\bibitem[{{Faherty} {et~al.}(2009){Faherty}, {Burgasser}, {Cruz}, {Shara}, {Walter}, \& {Gelino}}]{Faherty_BDKP}
{Faherty}, J.~K., {Burgasser}, A.~J., {Cruz}, K.~L., {et~al.} 2009, \aj, 137, 1, \dodoi{10.1088/0004-6256/137/1/1}

\bibitem[{{Faherty} {et~al.}(2021){Faherty}, {Bardalez Gagliuffi}, {Beichman}, {Burningham}, {Caselden}, {Eisenhardt}, {Gagne}, {Gelino}, {Gonzales}, {Kirkpatrick}, {Kuchner}, {Marocco}, {Meisner}, {Morley}, {Rothermich}, {Schneider}, {Vos}, \& {Whiteford}}]{Faherty_JWST}
{Faherty}, J.~K., {Bardalez Gagliuffi}, D.~C., {Beichman}, C.~A., {et~al.} 2021, {Explaining the Diversity of Cold Worlds}, JWST Proposal. Cycle 1, ID. \#2124

\bibitem[{{Fazio} {et~al.}(2004){Fazio}, {Hora}, {Allen}, {Ashby}, {Barmby}, {Deutsch}, {Huang}, {Kleiner}, {Marengo}, {Megeath}, {Melnick}, {Pahre}, {Patten}, {Polizotti}, {Smith}, {Taylor}, {Wang}, {Willner}, {Hoffmann}, {Pipher}, {Forrest}, {McMurty}, {McCreight}, {McKelvey}, {McMurray}, {Koch}, {Moseley}, {Arendt}, {Mentzell}, {Marx}, {Losch}, {Mayman}, {Eichhorn}, {Krebs}, {Jhabvala}, {Gezari}, {Fixsen}, {Flores}, {Shakoorzadeh}, {Jungo}, {Hakun}, {Workman}, {Karpati}, {Kichak}, {Whitley}, {Mann}, {Tollestrup}, {Eisenhardt}, {Stern}, {Gorjian}, {Bhattacharya}, {Carey}, {Nelson}, {Glaccum}, {Lacy}, {Lowrance}, {Laine}, {Reach}, {Stauffer}, {Surace}, {Wilson}, {Wright}, {Hoffman}, {Domingo}, \& {Cohen}}]{IRAC}
{Fazio}, G.~G., {Hora}, J.~L., {Allen}, L.~E., {et~al.} 2004, \apjs, 154, 10, \dodoi{10.1086/422843}

\bibitem[{{Filippazzo} {et~al.}(2015){Filippazzo}, {Rice}, {Faherty}, {Cruz}, {Van Gordon}, \& {Looper}}]{2015ApJ...810..158F}
{Filippazzo}, J.~C., {Rice}, E.~L., {Faherty}, J., {et~al.} 2015, \apj, 810, 158, \dodoi{10.1088/0004-637X/810/2/158}

\bibitem[{{Gagn{\'e}} {et~al.}(2020){Gagn{\'e}}, {David}, {Mamajek}, {Mann}, {Faherty}, \& {B{\'e}dard}}]{2020ApJ...903...96G}
{Gagn{\'e}}, J., {David}, T.~J., {Mamajek}, E.~E., {et~al.} 2020, \apj, 903, 96, \dodoi{10.3847/1538-4357/abb77e}

\bibitem[{Gagné {et~al.}(2018)Gagné, Mamajek, Malo, Riedel, Rodriguez, Lafrenière, Faherty, Roy-Loubier, Pueyo, Robin, \& Doyon}]{Gagné_2018}
Gagné, J., Mamajek, E.~E., Malo, L., {et~al.} 2018, The Astrophysical Journal, 856, 23, \dodoi{10.3847/1538-4357/aaae09}

\bibitem[{{Gaia Collaboration} {et~al.}(2018){Gaia Collaboration}, {Brown}, {Vallenari}, {Prusti}, {de Bruijne}, {Babusiaux}, {Bailer-Jones}, {Biermann}, {Evans}, {Eyer}, {Jansen}, {Jordi}, {Klioner}, {Lammers}, {Lindegren}, {Luri}, {Mignard}, {Panem}, {Pourbaix}, {Randich}, {Sartoretti}, {Siddiqui}, {Soubiran}, {van Leeuwen}, {Walton}, {Arenou}, {Bastian}, {Cropper}, {Drimmel}, {Katz}, {Lattanzi}, {Bakker}, {Cacciari}, {Casta{\~n}eda}, {Chaoul}, {Cheek}, {De Angeli}, {Fabricius}, {Guerra}, {Holl}, {Masana}, {Messineo}, {Mowlavi}, {Nienartowicz}, {Panuzzo}, {Portell}, {Riello}, {Seabroke}, {Tanga}, {Th{\'e}venin}, {Gracia-Abril}, {Comoretto}, {Garcia-Reinaldos}, {Teyssier}, {Altmann}, {Andrae}, {Audard}, {Bellas-Velidis}, {Benson}, {Berthier}, {Blomme}, {Burgess}, {Busso}, {Carry}, {Cellino}, {Clementini}, {Clotet}, {Creevey}, {Davidson}, {De Ridder}, {Delchambre}, {Dell'Oro}, {Ducourant}, {Fern{\'a}ndez-Hern{\'a}ndez}, {Fouesneau}, {Fr{\'e}mat}, {Galluccio}, {Garc{\'\i}a-Torres},
  {Gonz{\'a}lez-N{\'u}{\~n}ez}, {Gonz{\'a}lez-Vidal}, {Gosset}, {Guy}, {Halbwachs}, {Hambly}, {Harrison}, {Hern{\'a}ndez}, {Hestroffer}, {Hodgkin}, {Hutton}, {Jasniewicz}, {Jean-Antoine-Piccolo}, {Jordan}, {Korn}, {Krone-Martins}, {Lanzafame}, {Lebzelter}, {L{\"o}ffler}, {Manteiga}, {Marrese}, {Mart{\'\i}n-Fleitas}, {Moitinho}, {Mora}, {Muinonen}, {Osinde}, {Pancino}, {Pauwels}, {Petit}, {Recio-Blanco}, {Richards}, {Rimoldini}, {Robin}, {Sarro}, {Siopis}, {Smith}, {Sozzetti}, {S{\"u}veges}, {Torra}, {van Reeven}, {Abbas}, {Abreu Aramburu}, {Accart}, {Aerts}, {Altavilla}, {{\'A}lvarez}, {Alvarez}, {Alves}, {Anderson}, {Andrei}, {Anglada Varela}, {Antiche}, {Antoja}, {Arcay}, {Astraatmadja}, {Bach}, {Baker}, {Balaguer-N{\'u}{\~n}ez}, {Balm}, {Barache}, {Barata}, {Barbato}, {Barblan}, {Barklem}, {Barrado}, {Barros}, {Barstow}, {Bartholom{\'e} Mu{\~n}oz}, {Bassilana}, {Becciani}, {Bellazzini}, {Berihuete}, {Bertone}, {Bianchi}, {Bienaym{\'e}}, {Blanco-Cuaresma}, {Boch}, {Boeche}, {Bombrun}, {Borrachero},
  {Bossini}, {Bouquillon}, {Bourda}, {Bragaglia}, {Bramante}, {Breddels}, {Bressan}, {Brouillet}, {Br{\"u}semeister}, {Brugaletta}, {Bucciarelli}, {Burlacu}, {Busonero}, {Butkevich}, {Buzzi}, {Caffau}, {Cancelliere}, {Cannizzaro}, {Cantat-Gaudin}, {Carballo}, {Carlucci}, {Carrasco}, {Casamiquela}, {Castellani}, {Castro-Ginard}, {Charlot}, {Chemin}, {Chiavassa}, {Cocozza}, {Costigan}, {Cowell}, {Crifo}, {Crosta}, {Crowley}, {Cuypers}, {Dafonte}, {Damerdji}, {Dapergolas}, {David}, {David}, {de Laverny}, {De Luise}, {De March}, {de Martino}, {de Souza}, {de Torres}, {Debosscher}, {del Pozo}, {Delbo}, {Delgado}, {Delgado}, {Di Matteo}, {Diakite}, {Diener}, {Distefano}, {Dolding}, {Drazinos}, {Dur{\'a}n}, {Edvardsson}, {Enke}, {Eriksson}, {Esquej}, {Eynard Bontemps}, {Fabre}, {Fabrizio}, {Faigler}, {Falc{\~a}o}, {Farr{\`a}s Casas}, {Federici}, {Fedorets}, {Fernique}, {Figueras}, {Filippi}, {Findeisen}, {Fonti}, {Fraile}, {Fraser}, {Fr{\'e}zouls}, {Gai}, {Galleti}, {Garabato}, {Garc{\'\i}a-Sedano}, {Garofalo},
  {Garralda}, {Gavel}, {Gavras}, {Gerssen}, {Geyer}, {Giacobbe}, {Gilmore}, {Girona}, {Giuffrida}, {Glass}, {Gomes}, {Granvik}, {Gueguen}, {Guerrier}, {Guiraud}, {Guti{\'e}rrez-S{\'a}nchez}, {Haigron}, {Hatzidimitriou}, {Hauser}, {Haywood}, {Heiter}, {Helmi}, {Heu}, {Hilger}, {Hobbs}, {Hofmann}, {Holland}, {Huckle}, {Hypki}, {Icardi}, {Jan{\ss}en}, {Jevardat de Fombelle}, {Jonker}, {Juh{\'a}sz}, {Julbe}, {Karampelas}, {Kewley}, {Klar}, {Kochoska}, {Kohley}, {Kolenberg}, {Kontizas}, {Kontizas}, {Koposov}, {Kordopatis}, {Kostrzewa-Rutkowska}, {Koubsky}, {Lambert}, {Lanza}, {Lasne}, {Lavigne}, {Le Fustec}, {Le Poncin-Lafitte}, {Lebreton}, {Leccia}, {Leclerc}, {Lecoeur-Taibi}, {Lenhardt}, {Leroux}, {Liao}, {Licata}, {Lindstr{\o}m}, {Lister}, {Livanou}, {Lobel}, {L{\'o}pez}, {Managau}, {Mann}, {Mantelet}, {Marchal}, {Marchant}, {Marconi}, {Marinoni}, {Marschalk{\'o}}, {Marshall}, {Martino}, {Marton}, {Mary}, {Massari}, {Matijevi{\v{c}}}, {Mazeh}, {McMillan}, {Messina}, {Michalik}, {Millar}, {Molina}, {Molinaro},
  {Moln{\'a}r}, {Montegriffo}, {Mor}, {Morbidelli}, {Morel}, {Morris}, {Mulone}, {Muraveva}, {Musella}, {Nelemans}, {Nicastro}, {Noval}, {O'Mullane}, {Ord{\'e}novic}, {Ord{\'o}{\~n}ez-Blanco}, {Osborne}, {Pagani}, {Pagano}, {Pailler}, {Palacin}, {Palaversa}, {Panahi}, {Pawlak}, {Piersimoni}, {Pineau}, {Plachy}, {Plum}, {Poggio}, {Poujoulet}, {Pr{\v{s}}a}, {Pulone}, {Racero}, {Ragaini}, {Rambaux}, {Ramos-Lerate}, {Regibo}, {Reyl{\'e}}, {Riclet}, {Ripepi}, {Riva}, {Rivard}, {Rixon}, {Roegiers}, {Roelens}, {Romero-G{\'o}mez}, {Rowell}, {Royer}, {Ruiz-Dern}, {Sadowski}, {Sagrist{\`a} Sell{\'e}s}, {Sahlmann}, {Salgado}, {Salguero}, {Sanna}, {Santana-Ros}, {Sarasso}, {Savietto}, {Schultheis}, {Sciacca}, {Segol}, {Segovia}, {S{\'e}gransan}, {Shih}, {Siltala}, {Silva}, {Smart}, {Smith}, {Solano}, {Solitro}, {Sordo}, {Soria Nieto}, {Souchay}, {Spagna}, {Spoto}, {Stampa}, {Steele}, {Steidelm{\"u}ller}, {Stephenson}, {Stoev}, {Suess}, {Surdej}, {Szabados}, {Szegedi-Elek}, {Tapiador}, {Taris}, {Tauran}, {Taylor},
  {Teixeira}, {Terrett}, {Teyssandier}, {Thuillot}, {Titarenko}, {Torra Clotet}, {Turon}, {Ulla}, {Utrilla}, {Uzzi}, {Vaillant}, {Valentini}, {Valette}, {van Elteren}, {Van Hemelryck}, {van Leeuwen}, {Vaschetto}, {Vecchiato}, {Veljanoski}, {Viala}, {Vicente}, {Vogt}, {von Essen}, {Voss}, {Votruba}, {Voutsinas}, {Walmsley}, {Weiler}, {Wertz}, {Wevers}, {Wyrzykowski}, {Yoldas}, {{\v{Z}}erjal}, {Ziaeepour}, {Zorec}, {Zschocke}, {Zucker}, {Zurbach}, \& {Zwitter}}]{2018A&A...616A...1G}
{Gaia Collaboration}, {Brown}, A.~G.~A., {Vallenari}, A., {et~al.} 2018, \aap, 616, A1, \dodoi{10.1051/0004-6361/201833051}

\bibitem[{{Griffith} {et~al.}(2012){Griffith}, {Kirkpatrick}, {Eisenhardt}, {Gelino}, {Cushing}, {Benford}, {Blain}, {Bridge}, {Cohen}, {Cutri}, {Donoso}, {Jarrett}, {Lonsdale}, {Mace}, {Mainzer}, {Marsh}, {Padgett}, {Petty}, {Ressler}, {Skrutskie}, {Stanford}, {Stern}, {Tsai}, {Wright}, {Wu}, \& {Yan}}]{2012AJ....144..148G}
{Griffith}, R.~L., {Kirkpatrick}, J.~D., {Eisenhardt}, P. R.~M., {et~al.} 2012, \aj, 144, 148, \dodoi{10.1088/0004-6256/144/5/148}

\bibitem[{{Guillot}(1999)}]{Guillot_review}
{Guillot}, T. 1999, Science, 296, 72

\bibitem[{{Hambly} {et~al.}(2008){Hambly}, {Collins}, {Cross}, {Mann}, {Read}, {Sutorius}, {Bond}, {Bryant}, {Emerson}, {Lawrence}, {Rimoldini}, {Stewart}, {Williams}, {Adamson}, {Hirst}, {Dye}, \& {Warren}}]{wsa}
{Hambly}, N.~C., {Collins}, R.~S., {Cross}, N.~J.~G., {et~al.} 2008, \mnras, 384, 637, \dodoi{10.1111/j.1365-2966.2007.12700.x}

\bibitem[{Kirkpatrick {et~al.}(2010)Kirkpatrick, Looper, Burgasser, Schurr, Cutri, Cushing, Cruz, Sweet, Knapp, Barman, Bochanski, Roellig, McLean, McGovern, \& Rice}]{Kirkpatrick_2010}
Kirkpatrick, J.~D., Looper, D.~L., Burgasser, A.~J., {et~al.} 2010, The Astrophysical Journal Supplement Series, 190, 100, \dodoi{10.1088/0067-0049/190/1/100}

\bibitem[{Kirkpatrick {et~al.}(2011)Kirkpatrick, Cushing, Gelino, Griffith, Skrutskie, Marsh, Wright, Mainzer, Eisenhardt, McLean, Thompson, Bauer, Benford, Bridge, Lake, Petty, Stanford, Tsai, Bailey, Beichman, Bloom, Bochanski, Burgasser, Capak, Cruz, Hinz, Kartaltepe, Knox, Manohar, Masters, Morales-Calderón, Prato, Rodigas, Salvato, Schurr, Scoville, Simcoe, Stapelfeldt, Stern, Stock, \& Vacca}]{Kirkpatrick_2011}
Kirkpatrick, J.~D., Cushing, M.~C., Gelino, C.~R., {et~al.} 2011, The Astrophysical Journal Supplement Series, 197, 19, \dodoi{10.1088/0067-0049/197/2/19}

\bibitem[{Kirkpatrick {et~al.}(2012)Kirkpatrick, Gelino, Cushing, Mace, Griffith, Skrutskie, Marsh, Wright, Eisenhardt, McLean, Mainzer, Burgasser, Tinney, Parker, \& Salter}]{Kirkpatrick_2012}
Kirkpatrick, J.~D., Gelino, C.~R., Cushing, M.~C., {et~al.} 2012, The Astrophysical Journal, 753, 156, \dodoi{10.1088/0004-637X/753/2/156}

\bibitem[{Kirkpatrick {et~al.}(2019)Kirkpatrick, Martin, Smart, Cayago, Beichman, Marocco, Gelino, Faherty, Cushing, Schneider, Mace, Tinney, Wright, Lowrance, Ingalls, Vrba, Munn, Dahm, \& McLean}]{Kirkpatrick_2019}
Kirkpatrick, J.~D., Martin, E.~C., Smart, R.~L., {et~al.} 2019, The Astrophysical Journal Supplement Series, 240, 19, \dodoi{10.3847/1538-4365/aaf6af}

\bibitem[{Kirkpatrick {et~al.}(2021)Kirkpatrick, Gelino, Faherty, Meisner, Caselden, Schneider, Marocco, Cayago, Smart, Eisenhardt, Kuchner, Wright, Cushing, Allers, Gagliuffi, Burgasser, Gagné, Logsdon, Martin, Ingalls, Lowrance, Abrahams, Aganze, Gerasimov, Gonzales, Hsu, Kamraj, Kiman, Rees, Theissen, Ammar, Andersen, Beaulieu, Colin, Elachi, Goodman, Gramaize, Hamlet, Hong, Jonkeren, Khalil, Martin, Pendrill, Pumphrey, Rothermich, Sainio, Stenner, Tanner, Thévenot, Voloshin, Walla, Wędracki, \& Collaboration}]{Kirkpatrick_2021}
Kirkpatrick, J.~D., Gelino, C.~R., Faherty, J.~K., {et~al.} 2021, The Astrophysical Journal Supplement Series, 253, 7, \dodoi{10.3847/1538-4365/abd107}

\bibitem[{{Kirkpatrick} {et~al.}(2021){Kirkpatrick}, {Marocco}, {Caselden}, {Meisner}, {Faherty}, {Schneider}, {Kuchner}, {Casewell}, {Gelino}, {Cushing}, {Eisenhardt}, {Wright}, \& {Schurr}}]{The_Accident}
{Kirkpatrick}, J.~D., {Marocco}, F., {Caselden}, D., {et~al.} 2021, \apjl, 915, L6, \dodoi{10.3847/2041-8213/ac0437}

\bibitem[{{Kuchner} {et~al.}(2017){Kuchner}, {Faherty}, {Schneider}, {Meisner}, {Filippazzo}, {Gagn{\'e}}, {Trouille}, {Silverberg}, {Castro}, {Fletcher}, {Mokaev}, \& {Stajic}}]{Backyard_Worlds}
{Kuchner}, M.~J., {Faherty}, J.~K., {Schneider}, A.~C., {et~al.} 2017, \apjl, 841, L19, \dodoi{10.3847/2041-8213/aa7200}

\bibitem[{{Lacy} \& {Burrows}(2023)}]{2023ApJ...950....8L}
{Lacy}, B., \& {Burrows}, A. 2023, \apj, 950, 8, \dodoi{10.3847/1538-4357/acc8cb}

\bibitem[{{Leggett} {et~al.}(2019){Leggett}, {Apai}, {Burgasser}, {Cushing}, {Dupuy}, {Faherty}, {Gizis}, {Kirkpatrick}, {Marley}, {Morley}, {Schneider}, \& {Sousa-Silva}}]{Leggett_2019}
{Leggett}, S., {Apai}, D., {Burgasser}, A., {et~al.} 2019, \baas, 51, 95, \dodoi{10.48550/arXiv.1903.04686}

\bibitem[{{Leggett} {et~al.}(2017){Leggett}, {Tremblin}, {Esplin}, {Luhman}, \& {Morley}}]{Leggett_2017}
{Leggett}, S.~K., {Tremblin}, P., {Esplin}, T.~L., {Luhman}, K.~L., \& {Morley}, C.~V. 2017, \apj, 842, 118, \dodoi{10.3847/1538-4357/aa6fb5}

\bibitem[{{Leggett} {et~al.}(2021){Leggett}, {Tremblin}, {Phillips}, {Dupuy}, {Marley}, {Morley}, {Schneider}, {Caselden}, {Guillaume}, \& {Logsdon}}]{Leggett_2021}
{Leggett}, S.~K., {Tremblin}, P., {Phillips}, M.~W., {et~al.} 2021, \apj, 918, 11, \dodoi{10.3847/1538-4357/ac0cfe}

\bibitem[{{Luhman}(2014)}]{Luhman_2014}
{Luhman}, K.~L. 2014, \apjl, 786, L18, \dodoi{10.1088/2041-8205/786/2/L18}

\bibitem[{Mace {et~al.}(2013)Mace, Kirkpatrick, Cushing, Gelino, Griffith, Skrutskie, Marsh, Wright, Eisenhardt, Mclean, Thompson, Mix, Bailey, Beichman, Bloom, Burgasser, Fortney, Hinz, Knox, \& Stock}]{Mace_2013}
Mace, G., Kirkpatrick, J., Cushing, M., {et~al.} 2013, VizieR Online Data Catalog, 50006

\bibitem[{{Marley} {et~al.}(2021){Marley}, {Saumon}, {Visscher}, {Lupu}, {Freedman}, {Morley}, {Fortney}, {Seay}, {Smith}, {Teal}, \& {Wang}}]{2021ApJ...920...85M}
{Marley}, M.~S., {Saumon}, D., {Visscher}, C., {et~al.} 2021, \apj, 920, 85, \dodoi{10.3847/1538-4357/ac141d}

\bibitem[{Marocco {et~al.}(2021)Marocco, Eisenhardt, Fowler, Kirkpatrick, Meisner, Schlafly, Stanford, Garcia, Caselden, Cushing, Cutri, Faherty, Gelino, Gonzalez, Jarrett, Koontz, Mainzer, Marchese, Mobasher, Schlegel, Stern, Teplitz, \& Wright}]{Marocco_2021}
Marocco, F., Eisenhardt, P. R.~M., Fowler, J.~W., {et~al.} 2021, The Astrophysical Journal Supplement Series, 253, 8, \dodoi{10.3847/1538-4365/abd805}

\bibitem[{{Mart{\'\i}n} {et~al.}(2021){Mart{\'\i}n}, {Zhang}, {Esparza}, {Gracia}, {Rasilla}, {Masseron}, \& {Burgasser}}]{2021A&A...655L...3M}
{Mart{\'\i}n}, E.~L., {Zhang}, J.~Y., {Esparza}, P., {et~al.} 2021, \aap, 655, L3, \dodoi{10.1051/0004-6361/202142470}

\bibitem[{{Meisner} {et~al.}(2023){Meisner}, {Leggett}, {Logsdon}, {Schneider}, {Tremblin}, \& {Phillips}}]{Meisner_2023}
{Meisner}, A.~M., {Leggett}, S.~K., {Logsdon}, S.~E., {et~al.} 2023, \aj, 166, 57, \dodoi{10.3847/1538-3881/acdb68}

\bibitem[{{Meisner} {et~al.}(2020){Meisner}, {Caselden}, {Kirkpatrick}, {Marocco}, {Gelino}, {Cushing}, {Eisenhardt}, {Wright}, {Faherty}, {Koontz}, {Marchese}, {Khalil}, {Fowler}, \& {Schlafly}}]{CatWISE_Spitzer}
{Meisner}, A.~M., {Caselden}, D., {Kirkpatrick}, J.~D., {et~al.} 2020, \apj, 889, 74, \dodoi{10.3847/1538-4357/ab6215}

\bibitem[{Meisner {et~al.}(2021)Meisner, Schneider, Burgasser, Marocco, Line, Faherty, Kirkpatrick, Caselden, Kuchner, Gelino, Gagné, Theissen, Gerasimov, Aganze, chun Hsu, Wisniewski, Casewell, Gagliuffi, Logsdon, Eisenhardt, Allers, Debes, Allen, Andersen, Goodman, Gramaize, Martin, Sainio, Cushing, \& Collaboration}]{Meisner_2021}
Meisner, A.~M., Schneider, A.~C., Burgasser, A.~J., {et~al.} 2021, The Astrophysical Journal, 915, 120, \dodoi{10.3847/1538-4357/ac013c}

\bibitem[{{Miles} {et~al.}(2020){Miles}, {Skemer}, {Morley}, {Marley}, {Fortney}, {Allers}, {Faherty}, {Geballe}, {Visscher}, {Schneider}, {Lupu}, {Freedman}, \& {Bjoraker}}]{Miles_2020}
{Miles}, B.~E., {Skemer}, A. J.~I., {Morley}, C.~V., {et~al.} 2020, \aj, 160, 63, \dodoi{10.3847/1538-3881/ab9114}

\bibitem[{{Moranta} {et~al.}(2022){Moranta}, {Gagn{\'e}}, {Couture}, \& {Faherty}}]{2022ApJ...939...94M}
{Moranta}, L., {Gagn{\'e}}, J., {Couture}, D., \& {Faherty}, J.~K. 2022, \apj, 939, 94, \dodoi{10.3847/1538-4357/ac8c25}

\bibitem[{{Ricker} {et~al.}(2014){Ricker}, {Winn}, {Vanderspek}, {Latham}, {Bakos}, {Bean}, {Berta-Thompson}, {Brown}, {Buchhave}, {Butler}, {Butler}, {Chaplin}, {Charbonneau}, {Christensen-Dalsgaard}, {Clampin}, {Deming}, {Doty}, {De Lee}, {Dressing}, {Dunham}, {Endl}, {Fressin}, {Ge}, {Henning}, {Holman}, {Howard}, {Ida}, {Jenkins}, {Jernigan}, {Johnson}, {Kaltenegger}, {Kawai}, {Kjeldsen}, {Laughlin}, {Levine}, {Lin}, {Lissauer}, {MacQueen}, {Marcy}, {McCullough}, {Morton}, {Narita}, {Paegert}, {Palle}, {Pepe}, {Pepper}, {Quirrenbach}, {Rinehart}, {Sasselov}, {Sato}, {Seager}, {Sozzetti}, {Stassun}, {Sullivan}, {Szentgyorgyi}, {Torres}, {Udry}, \& {Villasenor}}]{TESS}
{Ricker}, G.~R., {Winn}, J.~N., {Vanderspek}, R., {et~al.} 2014, in Society of Photo-Optical Instrumentation Engineers (SPIE) Conference Series, Vol. 9143, Space Telescopes and Instrumentation 2014: Optical, Infrared, and Millimeter Wave, ed. J.~{Oschmann}, Jacobus~M., M.~{Clampin}, G.~G. {Fazio}, \& H.~A. {MacEwen}, 914320, \dodoi{10.1117/12.2063489}

\bibitem[{{Rieke}(2009)}]{Rieke_review}
{Rieke}, G.~H. 2009, Experimental Astronomy, 25, 125, \dodoi{10.1007/s10686-009-9148-7}

\bibitem[{Schneider {et~al.}(2015)Schneider, Cushing, Kirkpatrick, Gelino, Mace, Wright, Eisenhardt, Skrutskie, Griffith, \& Marsh}]{Schneider_2015}
Schneider, A.~C., Cushing, M.~C., Kirkpatrick, J.~D., {et~al.} 2015, The Astrophysical Journal, 804, 92, \dodoi{10.1088/0004-637X/804/2/92}

\bibitem[{Vacca {et~al.}(2003)Vacca, Cushing, \& Rayner}]{Vacca_2003}
Vacca, W.~D., Cushing, M.~C., \& Rayner, J.~T. 2003, Publications of the Astronomical Society of the Pacific, 115, 389, \dodoi{10.1086/346193}

\bibitem[{{Vos} {et~al.}(2017){Vos}, {Allers}, \& {Biller}}]{Vos_2017}
{Vos}, J.~M., {Allers}, K.~N., \& {Biller}, B.~A. 2017, \apj, 842, 78, \dodoi{10.3847/1538-4357/aa73cf}

\bibitem[{{Warren} {et~al.}(2007){Warren}, {Mortlock}, {Leggett}, {Pinfield}, {Homeier}, {Dye}, {Jameson}, {Lodieu}, {Lucas}, {Adamson}, {Allard}, {Barrado Y Navascu{\'e}s}, {Casali}, {Chiu}, {Hambly}, {Hewett}, {Hirst}, {Irwin}, {Lawrence}, {Liu}, {Mart{\'\i}n}, {Smart}, {Valdivielso}, \& {Venemans}}]{2007MNRAS.381.1400W}
{Warren}, S.~J., {Mortlock}, D.~J., {Leggett}, S.~K., {et~al.} 2007, \mnras, 381, 1400, \dodoi{10.1111/j.1365-2966.2007.12348.x}

\bibitem[{{Werner} {et~al.}(2004){Werner}, {Roellig}, {Low}, {Rieke}, {Rieke}, {Hoffmann}, {Young}, {Houck}, {Brandl}, {Fazio}, {Hora}, {Gehrz}, {Helou}, {Soifer}, {Stauffer}, {Keene}, {Eisenhardt}, {Gallagher}, {Gautier}, {Irace}, {Lawrence}, {Simmons}, {Van Cleve}, {Jura}, {Wright}, \& {Cruikshank}}]{Spitzer_Space_Telescope}
{Werner}, M.~W., {Roellig}, T.~L., {Low}, F.~J., {et~al.} 2004, \apjs, 154, 1, \dodoi{10.1086/422992}

\bibitem[{{Wilson} {et~al.}(2004){Wilson}, {Henderson}, {Herter}, {Matthews}, {Skrutskie}, {Adams}, {Moon}, {Smith}, {Gautier}, {Ressler}, {Soifer}, {Lin}, {Howard}, {LaMarr}, {Stolberg}, \& {Zink}}]{2004SPIE.5492.1295W}
{Wilson}, J.~C., {Henderson}, C.~P., {Herter}, T.~L., {et~al.} 2004, in Society of Photo-Optical Instrumentation Engineers (SPIE) Conference Series, Vol. 5492, Ground-based Instrumentation for Astronomy, ed. A.~F.~M. {Moorwood} \& M.~{Iye}, 1295--1305, \dodoi{10.1117/12.550925}

\bibitem[{{Wright} {et~al.}(2010){Wright}, {Eisenhardt}, {Mainzer}, {Ressler}, {Cutri}, {Jarrett}, {Kirkpatrick}, {Padgett}, {McMillan}, {Skrutskie}, {Stanford}, {Cohen}, {Walker}, {Mather}, {Leisawitz}, {Gautier}, {McLean}, {Benford}, {Lonsdale}, {Blain}, {Mendez}, {Irace}, {Duval}, {Liu}, {Royer}, {Heinrichsen}, {Howard}, {Shannon}, {Kendall}, {Walsh}, {Larsen}, {Cardon}, {Schick}, {Schwalm}, {Abid}, {Fabinsky}, {Naes}, \& {Tsai}}]{Wright_2010}
{Wright}, E.~L., {Eisenhardt}, P. R.~M., {Mainzer}, A.~K., {et~al.} 2010, \aj, 140, 1868, \dodoi{10.1088/0004-6256/140/6/1868}

\bibitem[{{Zhang} {et~al.}(2019){Zhang}, {Burgasser}, {G{\'a}lvez-Ortiz}, {Lodieu}, {Zapatero Osorio}, {Pinfield}, \& {Allard}}]{Zhang_2019}
{Zhang}, Z.~H., {Burgasser}, A.~J., {G{\'a}lvez-Ortiz}, M.~C., {et~al.} 2019, \mnras, 486, 1260, \dodoi{10.1093/mnras/stz777}

\end{thebibliography}
\bibliographystyle{aasjournal}

\end{document}